\documentstyle[emulateapj]{article}

\def\degr{\hbox{$^\circ$}}

\slugcomment{To appear in Astronomical Journal}

\lefthead{Oliveira et al.}

\righthead{Candidate Tidal Dwarf Galaxies in H92}

\input epsf.sty
\begin{document}

\title{Candidate tidal dwarf galaxies associated with the Stephan's Quintet}
 
\author{C. Mendes de Oliveira \altaffilmark{1},
H. Plana \altaffilmark{2,3}, P. Amram \altaffilmark{3}, C. Balkowski 
\altaffilmark{4}, M. Bolte \altaffilmark{5}} 
 
\altaffiltext{1}{ Instituto Astron\^omico e Geof\'{\i}sico (IAG),
      Av Miguel Stefano 4200 CEP: 04301-904 S\~ao Paulo, Brazil}
 
\altaffiltext{2}{Observatorio Astronomico Nacional, UNAM, Apartado Postal 877, 
22800, Ensenada, BC, M\'exico}
 
\altaffiltext{3}{ Laboratoire d'Astrophysique de Marseille, Observatoire
de Marseille, 2 Place Leverrier, 13248 Marseille Cedex 04 France}

\altaffiltext{4}{Observatoire de Paris, DAEC and UMR 8631, CNRS and 
Universite Paris 7, 5 Place Jules Janssen, F-92195 Meudon Cedex, France}

\altaffiltext{5}{ Lick Observatory, Board of Studies in Astronomy and Astrophysics, University of California, Santa Cruz, California 95064}

\begin{abstract}

  We present kinematic and photometric evidence for the presence of
seven candidate tidal dwarf galaxies in Stephan's quintet.
The central regions of the two most probable parent galaxies, N7319
and N7318B, contain little or no gas whereas the intragroup medium, and
particularly the optical tails that seem to be associated with N7318B
are rich in cold and ionized gas.

   Two tidal-dwarf candidates may be located at the edge of a tidal tail,
one within a tail and for four others there is no obvious stellar/gaseous
bridge between them and the parent galaxy.  Two of the candidates are
associated with HI clouds, one of which is, in addition, associated with a CO
cloud.  All seven regions have low continuum fluxes and high H$\alpha$
luminosity densities (F(H$\alpha$) = 1 -- 60 $\times$ 10$^{-14}$ erg
s$^{-1}$ cm$^{-2}$).  Their magnitudes (M$_B =$ --16.1 to --12.6), sizes
($\sim$ 3.5 h$_{75}^{-1}$ kpc), colors (typically $B-R = 0.7$) and gas
velocity gradients ($\sim$ 8 -- 26 h$_{75}$ km s$^{-1}$ kpc$^{-1}$) are typical
for tidal dwarf galaxies.  In addition the ratios between their star formation
rates determined from H$\alpha$ and from the B band luminosity 
are typical of other
tidal dwarf galaxies. The masses of the tidal dwarf galaxies in
Stephan's quintet range from $\sim$ 2 $\times$ 10$^8$ to  
10$^{10}$ M$_\odot$ and the median value for their inferred mass-to-light
ratios is 7 M$_\odot$/L$_\odot$.

   At least two of the systems may survive possible
``fall-backs'' or disruption by the parent galaxies and may already be
or turn into self-gravitating dwarf galaxies, new members of the group.

\end{abstract}


\section{Introduction}

   Optical and HI studies of interacting galaxies (e.g. Hibbard
\& van Gorkom 1996, Duc \& Mirabel 1998) have shown that dwarf
galaxies may be produced during galactic collisions.  Duc \& Mirabel
(1998) have presented an especially convincing case of NGC 5291, where
more than ten star-forming high-metallicity dwarf galaxies may have
been formed in a recent merger. These newly formed tidal-dwarf
galaxies may be good sites to study galaxy formation in the 
nearby universe.

  Although the concept of the formation of self-gravitating objects in
tidal tails is an old one, considered by Zwicky as early as in 1956,
it is only recently that these objects have received more attention
and have been searched for systematically.  HI observations have so far
been the principal means used in searches of tidal dwarf galaxies since
these are usually gas-rich systems. HI maps can reach large radii but
have  the disadvantage of being (mostly) of too poor spatial resolution
for a direct investigation of the velocity field of the spatially-small
and low-velocity-dispersion objects (typically 5-20 arcsec in extent,
and velocity gradients of 2 -- 20 km s$^{-1}$ arcsec$^{-1}$; Duc, 1995).  
The study of
the kinematics of the {\it ionized} gas of a galaxy, when it is present,
with high spatial and spectral resolution velocity maps, can alternatively
provide a useful tool to probe the nature of the candidate dwarf galaxies
at small scales, complementing the lower-resolution radio studies.

  This paper presents new data on the kinematics of the ionized gas
for seven star-forming tidal dwarf galaxy candidates and many other giant HII
regions of Stephan's quintet. Combining Fabry Perot H$\alpha$ maps with
deep B and R images, we study the photometric and kinematic properties
of 23 emission-line regions of the group.

   Stephan's Quintet (Arp 319, VV288, HCG 92)
has been
the subject of studies in x-rays, radio continuum, HI and H$\alpha$.
Most of its gaseous material is not concentrated around the
bright galaxies but in the intragroup medium, suggesting that
collisions among the group members may have taken place.  Arp (1973)
was the first to obtain an H$\alpha$ image of the group and to identify
HII regions around galaxies N7318A,B, N7319 and N7320. As part of an
imaging survey of compact groups, in the R band, Hunsberger et al.
(1996) identified 27 tidal dwarf galaxy candidates possibly associated
with N7318A,B and N7319.  One of these regions was observed by Xu et
al. (1999) to be a bright infrared source, when observed with the ISO
satellite (their source A).  Xu et al. (1999) also presented H$\alpha$
+ [NII] images of the group highlighting the spatial overlap of the two
emission clouds of ionized gas in the intragroup medium centered around
5700 km s$^{-1}$ and 6700 km s$^{-1}$. A similar result was found by Plana
et al.  (1999) based on Fabry-Perot H$\alpha$ maps.

 An interesting possible scenario for the interaction history of
Stephan's quintet has been put forward by Shostak et al. (1984) and
refined by Moles et al.  (1997, 1998). The scenario is based on the analysis of
optical images of the quintet, spectroscopy of a number of systems (Moles
et al. 1998) and HI maps which show three velocity components associated
with the group, at $\sim$ 5700 km s$^{-1}$, 6000 km s$^{-1}$ and 6700
km s$^{-1}$ (Balkowski et al. 1973, Allen \& Sullivan 1980, Shostak
et al.  1984).  In their scenario, one or more major collisions occurred
between N7320C and N7319 (at velocities $\sim$ 6000 km s$^{-1}$ and $\sim$ 6700
km s$^{-1}$ respectively), which may have resulted in the removal of the
HI gas from the inner parts of N7319.  More recently, Stephan's quintet
would have received a new member, N7318B (it has a radial velocity of
5774 km s$^{-1}$ while the other core-group members, N7317, N7318A and
N7319, have a mean velocity of 6664 km s$^{-1}$, Hickson et al. 1992).
N7318B would be entering the group for the first time and it would be
colliding with the intragroup medium (Moles et al. 1997).  We identify a
number of possible tidal dwarf galaxy candidates most probably formed as
a result of the ongoing collision that involves N7318B.
Only one of the candidates studied here may be associated with
the collision N7320C--N7319 and hence the HI cloud removed from the center of
N7319. An alternative possibility is that the latter originated
from N7318A (see Sections 4.7 and 4.8).

 All the regions we
measure in this work 
have been previously identified as HII regions by Arp (1973).
A number of them have also been listed by Hunsberger et al. (1996) 
as tidal dwarf galaxy candidates and Gallagher et al. (2000, using
HST images) as massive cluster candidates.  However, these studies
did not have velocity information.  Radial velocity information for
some of the regions is summarized in Table 1 of Plana et al. (1999).
Determination of the velocity gradients within the regions is the main
contribution of the present work.

    The organization of this paper is as follows. Section 2 describes
the observations and reduction procedure. In Section 3, we present
the new photometric and kinematic results. In Section 4, we discuss
the results and section 5 summarizes the paper. Throughout this paper
h$_{75}$ is the Hubble constant in units of H$_0$ = 75 km s$^{-1}$
Mpc$^{-1}$.  We assume that all the regions are at the same distance,
and the differences in cz are just kinematical.  For the figures and
the determination of the star-formation rates we assumed a distance to
the group of 80 Mpc, for an average velocity of 6000 km s$^{-1}$ and a
Hubble constant of 75 km s$^{-1}$ Mpc$^{-2}$.

\section{Observations and Data Reduction}

\subsection{Fabry Perot data}

 The observations were carried out with a scanning Fabry-Perot
instrument mounted on the Canada-France-Hawaii 3.6m telescope (CFHT)
and the Russian SAO 6m telescopes.  The observations and characteristics
of the set-ups are summarized in Table 1. During the CFHT run we mostly
detected sources with radial velocities V$_R$ $\sim$ 5700 km s$^{-1}$,
since we used a narrow-band filter that cut the higher velocities. We,
therefore, did not detect the 6700 km s$^{-1}$ velocity component of
the group.  The filter used during the observations at the Russian SAO 6m
telescope allowed the detection of a much broader range of wavelengths
and both velocity components were detected. However, the data are of
much lower signal-to-noise and spectral and spatial resolution than the
CFHT data.

%
%

\begin{deluxetable}{llll}
\small
\tablenum{1}
\tablecolumns{4}
\tablewidth{0pt}
\tablecaption{Journal of Perot-Fabry observations \label{tbl-1}}
\tablehead{
\colhead{} & \colhead{HCG 92 or Stephan's quintet} & \colhead{}}
\startdata
Observations & Telescope & CFHT 3.6m & SAO 6m \\ 
	     & Equipment & MOS/FP @ Cassegrain & CIGALE  @ Primary 
\\ 
	     & Date & August, 23th 1996 & August, 5th 1991 \\ 
	     & Seeing & $<$ 1" & $\sim$ 1.2" \\ 
Interference Filter &   Central Wavelength & 6697 \AA \tablenotemark{1} & 6726 
\AA  \tablenotemark{2} \\ 
		    &   FWHM & 17 \AA  \tablenotemark{1} & 50 \AA 
\tablenotemark{2} \\ 
		    &   Transmission (Maximum) & 0.7 \tablenotemark{1} & 0.55 
\tablenotemark{2} \\ 
Calibration & Neon Comparison light & $\lambda$ 6598.95 \AA & $\lambda$ 6598.95 
\AA \\ 
Perot--Fabry & Interference Order & 1162 @ 6562.78 \AA & 796 @ 6562.78 \AA \\ 
		 & Free Spectral Range at H$\alpha$ & 265 km s$^{-1}$ & 380 km 
s$^{-1}$ \\ 
		 & Finesse at H$\alpha$ & 12 & 12\\ 
                 & Spectral resolution at H$\alpha$ & 27344 at the sample step & 
 18750 \\
Sampling & Number of Scanning Steps & 24 & 24 \\ 
	 & Sampling Step & 0.24 \AA\ (11 km s$^{-1}$) & 0.35 \AA\ (16 km 
s$^{-1}$)\\ 
	 & Total Field & 440''$\times $440'' (512$\times $512 px$^2$) & 
 170''$\times $170'' (256$\times $256 px$^2$) \\ 
	 & Pixel Size & 0.86'' & 0.67''  \\ 
Detector & & STIS 2 CCD & IPCS  \\ 
Exposures times & Total exposure time & 1.2 hours & 0.8 hour \\ 
		& Each scanning exposure time & 180 s & 20 s per channel \\ 
		& Total exposure time per channel & 180 s & 120 s\\

\enddata
\tablenotetext{1} {For a mean beam inclination of 2.7\degr }
\tablenotetext{2} {For a mean beam inclination of 5.3\degr }

\end{deluxetable}

   Reduction of the data was performed using the CIGALE/ADHOC software
(http://www-obs.cnrs-mrs.fr/adhoc/adhoc.html).  The data reduction
procedure has been extensively described in Amram et al. (1995, 1996),
Plana et al.  (1998) and references therein.

   Wavelength calibrations were obtained by scanning the narrow Ne 6599
\AA\ line during the observations.  The relative velocities with
respect to the systemic velocities are very accurate, with an error of a
fraction of a channel width (${\rm <3 \, km s^{-1}}$) over the whole
field.

   The signal measured along the scanning sequence was separated into
two parts: (1) a constant level produced by the continuum light
in a narrow passband around H$\alpha $ (continuum map), and (2) a
varying part produced by the H$\alpha $ line (monochromatic map).  The
continuum level was taken to be the mean of the three faintest
channels. The monochromatic map was obtained by integrating the
monochromatic profile in each pixel.  

   The CFHT data were roughly calibrated in flux using a common galaxy
also observed in a previous run when a flux calibrator was observed
(Amram et al. 1998).  We estimate an error of 30\% in the flux
calibration.  The H$\alpha$ profiles were measured to a minimum flux
density of 4.9 $\times$ 10$^{-17}$ erg s$^{-1}$cm$^{-2}$
arcsec$^{-2}$. We were not able to flux calibrate the data 
obtained at the SAO 6m telescope for lack of calibrators. 

\subsection {Imaging}

   Images of Stephan's quintet in B (5 x 900s) and R (6 x 350s) were
obtained with the CFHT and the Subarcsec
Imaging instrument.  The values of seeing on the final images were 0.8
and 0.6 arcsec respectively.

   Preprocessing of all frames through flat-fielding was carried out using
IRAF tasks. The individual frames were registered and combined with a
clipping scheme that eliminated one lowest and one highest pixel values
in each registered stack to exclude cosmic rays, hot pixels and low
pixels. 

  The program SExtractor (Bertin \& Arnouts 1996) was used to obtain
the photometry of all objects in the B and R images with a sky ``meshsize''
of 128 pix.  ``Automatic--aperture magnitudes'' were obtained, which
give total magnitudes for the objects (based on the Kron's, 1980,
first-moment algorithm).

 Calibration to the standard system for the B and R band observations
was done via observations of standard stars from the list of Landolt
(1992) and updated magnitudes for the KPNO consortium fields in M92 and
NGC 7006.  The transformations from instrumental to standard systems
were made using the following:  $R$ = $r +$ 0.015($B-R$) $+$ $31.37$
and $B$ = $b +$ 0.047($B-R$) $+$ $31.27$. The final combined images of
H92 can be obtained upon request.  A B--R map was obtained after the
images were calibrated and their seeing profiles were matched (shown in
Fig. 1a).

\section{Results}

\subsection{Magnitudes and colors of the emission-line regions}

   The H$\alpha$ contours superimposed on the B--R image of Stephan's
quintet is shown in Fig. 1a.  The 23 emitting regions detected by Plana
et al. (1999) are marked. Dark is blue and white is red. The letters
``L'' and ``H'' are used to identify the systems for which we have an
unambiguous determination of their radial velocities. The ones marked with
``L'' belong to the low-velocity system associated with N7318B, with
velocities of 5700--6000 km s$^{-1}$ and the only system marked with
``H'', region 6, is the high-velocity system probably associated with
N7319, at a velocity of $\sim$ 6700 km s$^{-1}$.  

   In Fig. 1b the corresponding R image of the group (after subtraction
of a heavily smoothed sky) is shown. The optical blobs corresponding to
the H$\alpha$ emitting regions are marked in this figure (e.g. regions
1A and 1B  correspond to the B and R counterparts of region 1 in
H$\alpha$, see Fig. 1a).  All regions detected in H$\alpha$ have
counterparts on the B and R image.  Region 17 is faint, it is
affected by dust, and it is too close to the galaxy
N7318A. Although it is not evident given the contrast of Fig. 1b, it
does have a counterpart on the B and R images.

  The B--R image of the group
shown in Fig. 1a reveals some interesting features.  The galaxy N7319
may contain several dust spots (white regions in the figure, which are
those with the reddest values of B--R), particularly the one to the
north of the galaxy corresponds to strong CO emission (Gao \& Xu 2000).
We also note that the center of galaxy N7318A has a dust lane with a very
peculiar shape, suggesting that this galaxy may have taken part in
the interaction history of the group (not contemplated in the scenario
described by Moles et al. 1997). The complicated dust lane structure of
N7318A is much more impressive in the HST image of the group obtained
by Charlton et al. (2000).  Further evidence that galaxy N7318A may be
in interaction is given by Fig. 1c, which is discussed in Section 4.7.

%
%

\begin{deluxetable}{lcccccccc}
\small
\tablenum{2}
\tablecolumns{9}
\tablewidth{0pt}
\tablecaption{Properties of the star-forming regions of Stephan's quintet  \label{tbl-2}}
\tablehead{\colhead{ID} & \colhead{B} & \colhead{M$_B$} & \colhead{(B--R)$_0$}&  \colhead{Cont.}& \colhead{F(H$_\alpha$)\tablenotemark{a}} & \colhead{SFR(H$\alpha$)\tablenotemark{b}} & \colhead{SFR(L$_B$)\tablenotemark{b}} & \colhead{a $\times$ b} \\
\colhead{}& \colhead{(mag)}& \colhead{(mag)} &\colhead{(mag)}& \colhead{(\%)} & \colhead{ $\times$ 10$^{-14}$} &  \colhead{} & \colhead{$\times$ 10$^{-3}$} & \colhead{(arcsec)} } 
\startdata
 1 &      20.7/21.9  &    --14.2   &   0.76 & 3.1  &  1.5 &    0.23 &   2.2
&    9.0 x 8.4        \nl
 2\tablenotemark{c} &      20.6  &    --14.0   &   0.53 & 2.3  &  28.7 &    4.26  &
12.2 &  34 $\times$ 17        \nl
 3\tablenotemark{c} &      20.3/20.9  &    --14.7   &   0.55 & 3.3  &  --    &        -- &
 &  \nl
 4\tablenotemark{c} &      21.2  &    --13.4   &   0.67 & 3.2  &  --    &        -- &
 &  \nl
 5\tablenotemark{c} &      19.6/20.7  &    --15.3   &   0.47 & 3.9  &  --    &        -- &
 &  \nl
 6\tablenotemark{c} &      19.5  &    --15.0   &   1.32 & 3.6  &  --  &    -- &
4.5 & 16.0 x 13.0 \nl
 7 &      21.5  &    --13.0   &   0.75 & 3.2  &  0.3 &    0.04 &  0.7 
& 2.6 x 3.4 \nl
 8* &      21.4/22.3/21.2  &    --14.2   &   0.79 & 3.0  &  2.5 &    0.38 &
2.2  & 13.7 x 7.5 \nl
 9 &      23.2  &    --11.4   &   1.00 & --   &  2.5 &    0.38 &   0.2
& 7.4 x 4.6 \nl
10 &      20.1  &    --14.5   &   0.92 & 6.3  &  0.7 &    0.09 &  2.9 
& 3.1 x 3.1 \nl
11 &      20.8  &    --13.7   &   0.58 & 4.8  &  0.7 &    0.09 &   1.4
& 4.7 x 3.1 \nl
12 &      19.6  &    --4.9   &   0.43 & 4.6  &  0.7 &    0.10 &  4.1 
& 8.3 x 4.8 \nl
13 &      21.5  &    --13.0   &   0.77 & --   &  2.2 &    0.32 &  0.7 
& 8.6 x 5.6 \nl
14\tablenotemark{d} &      19.3/21.2  &    --15.4   &   0.94 & 6.3  &  8.1 &    1.20 & 6.5 
& 9.2 x 6.7 \nl
15 &      20.8/20.4  &    --14.7   &   1.00 & 4.7  &  9.6 &    1.42 &  3.4 
& 10.8 x 9.3 \nl
16 &    21.1/21.6/22.0/21.5  &    --14.3   &   0.86 & 5.4  &  3.2 &    0.46 &  
2.4 & 3.3 x 3.1  \nl
17 &          --  &        --   &       -- &  --  &  2.7 &    0.40 &
--    & 5.4 x 3.5 \nl
18 &      19.7/21.0  &    --15.1   &   0.56 & 4.0  &  13.4 &    2.00 &
4.9 & 11.9 x 9.0   \nl
19 &      20.3/21.5  &    --14.6   &   0.71 & 2.9  &  9.0 &    1.35&  3.1 
& 15.0 x 7.2 \nl
20\tablenotemark{c} &      21.9  &    --12.6   &   0.76 & 0.3  &  3.0 &    0.45 &
0.5 & 10.5 x 8.0 \nl
21\tablenotemark{c} &      20.1  &    --14.4   &   0.34 & 2.0  &  4.9 &    0.73 &
2.6 & 11.3 x 10.5 \nl
22\tablenotemark{c} &      21.0/21.7/22.2  &    --14.2   &   0.42 & 1.0  &  4.2 &    0.63 &
2.2 & 11.8 x 9.7 \nl
23\tablenotemark{c} &      21.5  &    --13.0   &   0.63 & 0.2  &  1.3 &    0.20&  0.7 
& 8.2 x 8.7 \nl
\enddata
\tablenotetext{a}{the units are erg s$^{-1}$ cm$^{-2}$}
\tablenotetext{b}{the units are M$_{\odot}$ yr$^{-1}$ L$_{\odot}^{-1}$} 
\tablenotetext{c} {objects classified as tidal dwarf galaxies}
\tablenotetext{d} {the magnitude and color of this region are contaminated by 
light from a red background galaxy}
\end{deluxetable}

   Table 2 shows the main photometric parameters for all the 23
H$\alpha$-emitting regions shown in Figs. 1a and 1b (with exception of
a few entries for region 17, for the reason explained previously). Total B
magnitudes (``automatic-aperture magnitudes'' given by SExtractor) and
B--R colors are listed in columns 2 and 3.  When one H$\alpha$ region
corresponds to more than one identification on the B and R images we
list the apparent B magnitudes of the different blobs in a
single entry, separated by slashes (magnitudes and colors are corrected
for Galactic extinction only).  The B--R color was measured inside an
aperture of 1.0 arcsec radius centered on each blob.  The total absolute
magnitudes given in column 3 correspond to the sums of the individual
luminosities (e.g. the absolute magnitude for region 8, see Fig. 1a,
corresponds to the sum of regions 8A, 8B and 8C, see Fig. 1b). The 
ID's marked with a ``c'' on Table 2 correspond to the candidate
tidal dwarf galaxies (see section 3.3). The magnitude and color of 
region 14A is contaminated by light from 
a red background galaxy that overlaps with this
region. There is an object to the northwest of region 2 which also seems to 
be a background object, given its red color. 

  We measure a very low continuum level (on the Fabry-Perot map)
for several of the H$\alpha$-emitting regions.
Since we were not able to perform calibration in flux of the continuum
map (for lack of calibrators), we determined the ratio of the total
flux contained in each blob to the flux within an arbitrary aperture of
radius 1.8" centered on galaxy N7318B. This fraction (in percentage) is
shown in column 5 of Table 2.  This procedure is intended to show, in a
relative way, which regions are the ones with brightest and faintest
continuum emission.  The regions with the lowest fluxes are regions 20
and 23. They have $\sim$ 10 times less flux in the continuum than the
median for the remaining regions. Regions 9, 13 and 17 could not be
measured due to contamination from the bright galaxies N7318A,B.

   We show color gradients for each emission-line blob in Fig. 2.
The (B--R)$_0$ (corrected for galactic extinction)
profiles in general become bluer toward each HII region, despite the fact
that the R-band flux is contaminated by the H$\alpha$-line emission.
We can see from column 4 of Table 2 that all but one emitting region
detected in H$\alpha$ are blue (the median value of the color is 0.7).
The exception is region 6, with a B--R color of 1.3.

\subsection{The kinematics and the nature of the emission-line regions}

    Plana et al. (1999) detected 23 H$\alpha$ emitting regions around
the N7318A/B system. For seven of them (six systems and one complex with
four regions) a velocity field is derived and shown in Figs. 3a to 3g.
These correspond to the objects which show rotation.  The velocity fields
were used to derive the rotation curves shown in Fig.  4.  In all cases
the velocities were measured within $\pm35\degr$ of the major axis in
the plane of the sky, except for regions 2--5 and 6, where a cone of
$\pm40\degr$ was used instead.  The rotation curves were obtained using
the position angle and inclination listed in columns 2 and 3 of Table 3.
The position angle of the major axis (given in column 2) is taken to
be the axis of symmetry of the main kinematic body of each galaxy.
The center of the velocity field is whenever possible chosen to be a
point along the major axis which makes the rotation curve symmetric,
with similar amplitudes for the receding and approaching sides (this is,
however, not always possible, e.g. region 6).  The values of inclination
with respect to the plane of the sky (given in column 3) are determined
from the velocity fields of the candidates.  Specifically we used a least
square's fit program which gives the best value for the inclination fixing
the position angle, the central systemic velocity and the kinematic
center and by fitting a classical rotation velocity model for data points
inside crowns at increasingly growing distances from the center.

  In the top-left panels of Figs. 3a--g we show the H$\alpha$ contours
determined from the CFHT Fabry-Perot maps, superimposed on the deep
B-band image of the galaxy (see also Plana et al. 1999).  For region 6
the H$\alpha$ map is derived from the SAO 6m data (since its velocity was
outside the filter range for the CFHT observations). The lower-left panels
show the continuum-subtracted emission-line profiles.  These have been
smoothed spectrally by a gaussian function having a FWHM of three channels
(or 35 km s$^{-1}$) and spatially (3x3 pixels), except for Fig. 3a (see
below).  Each pixel represents 0.86" on the sky. Although the pixel size
for the image of region 6 was originally 0.67", it was degraded to 0.86"
to match the CFHT data.  In the right panels we show the velocity fields
of the galaxies superimposed on the corresponding monochromatic H$\alpha$
images. We mark the absolute value for each isovelocity for regions 2--5,
6 and 21, but for regions 8, 20, 22 and 23, only relative velocities
are marked, since we do not know the systemic velocities of these regions.

  We classify systems 2--5, 6, 8, 20, 21, 22 and 23 as tidal dwarf
galaxy candidates on the basis of the clear signature of rotation,
above a cut-off value of 8.0 h$_{75}$ km s$^{-1}$ kpc$^{-1}$). Examples
of non-rotating structures, possible associations of giant HII regions,
are regions 9, 10, 11, 14, 15, 18 and 19.

  We see continuous profiles between regions 9 to 11 (the data
have S/N $>$ 20 in between the emission blobs) showing that these are
connected. These regions are most probably attached to the southern tail
of N7318B. No velocity gradients are detected within them.  Regions 14
and 15 are attached to one or more tidal tails (these regions correspond
to double velocity components, Plana et al. 1999). 
Even if these complexes have H$\alpha$
fluxes, sizes and colors consistent with their being tidal dwarf galaxies
(like the ones studied by Duc 1995), we detect no velocity gradient
within the objects.  In the same way, regions 18 and 19 also show
bright H$\alpha$ emission with no velocity gradients.  These are very
close to N7318A but, as indicated by their
radial velocities ($\sim$ 5700 km s$^{-1}$, Plana et al. 1999), 
they may belong to N7318B.
Regions 18 and 19 do not have any obvious contact to a tidal arm.

  Region 1 is isolated.  We detected no velocity gradient within the
region but the S/N of the data may be too low to allow a gradient measurement.

   We summarize below the properties of each object that we classify as
candidate tidal dwarf galaxies.

\subsection {Properties of the candidate tidal dwarf galaxies}

{\bf Regions 2--5:} The complex formed by regions 2--5 has a 
radial velocity of 6020 km s$^{-1}$ (velocity obtained by 
Moles et al. 1998 for region 2)
and it shows an H$\alpha$
diameter (full width at zero intensity) of $\sim$ 34" 
(corresponding to a radius of R=6.5 h$_{75}^{-1}$ kpc). In order to increase the
S/N ratio we performed a spatial smoothing of 3x3 pixels in the center
and 5x5 pixels outside the inner region. Since this complex is large,
smoothing does not affect the results. We can see on the lower-left
panel of Fig. 3a that the complex formed by regions 2--5 is connected
by continuous velocity profiles.  Although with a much lower S/N, the
effect is also measurable in the unbinned data.  This suggested to us
that these regions may form one system, perhaps within a tidal tail.
Fig. 4a shows the corresponding rotation curve.  The approaching and
receding sides of the rotation curve differ in shape and amplitude. The
curve is increasing regularly, reaching the average maximum velocity of
50 km s$^{-1}$. As the monochromatic map shows, we have at least four
emitting regions and therefore the velocity field and rotation curve may
be averages of internal motions of these various regions.  This could
explain, at least partially, the disturbances of the velocity field.
The R image in Fig. 1b shows at least six bright condensations. The system
as a whole has no defined center. The conditions needed for the complex of
regions 2--5 to become an independent galaxy are discussed in Sections 4.1
and 4.6. We
conclude that there is little chance that this region will survive as a
single object.  It could, however, form several condensations that could
then be stable.  We point out that there is an HI cloud encompassing
the region with velocity similar to the optical velocity measured from
the H$\alpha$ emission ($\sim$ 6000 km s$^{-1}$, Williams et al. 1999).

{\bf Region 6:}   Region 6 has a radial velocity of 6680 km s$^{-1}$
(Moles et al. 1998).  It shows a disturbed and fairly extended velocity
curve out to a radius of R=3.5 h$_{75}^{-1}$ kpc. The profiles shown in
the lower-left panel of Fig. 3b indicate a larger velocity dispersion
(i.e.  they have larger widths) than for the other regions ($\sim$ 115 km
s$^{-1}$ compared to 80 km s$^{-1}$ for the others). A larger dispersion
is expected when there is, for example, a second component blended with
the main velocity component.  However, in the case of region 6, the
components are not resolved and any attempt to decompose them would be
arbitrary.  As can be seen in Fig. 3b, some of the profiles do show
double peaks but no consistent solution for a moving second component
could be found, perhaps due to low S/N.  The rotation curve presented
in Fig. 4b is peculiar.  There is complete disagreement between the
amplitudes of the receding and the approaching sides.  The approaching
side has approximately constant velocity at $\sim$ 10 km s$^{-1}$ while
for the receding side the velocity reaches 90 km s$^{-1}$.  The maximum
rotational velocity listed in Table 4 is taken to be the average between
both sides. The mass estimate obtained from this number is therefore
very uncertain. Williams et al. (1999) have detected an HI cloud at
the position of Region 6 with similar velocity to that measured from the
H$\alpha$ emission ($\sim$ 6700 km s$^{-1}$). It was also detected 
in CO by Gao \& Xu (2000). In addition, this region
was detected with ISO by Xu et al. (1999) and it was found to be a very
bright source in the infrared. This could therefore be a peculiar object,
given its infrared luminosity and its disturbed rotation curve.

{\bf Region 8:}  It shows an extent of R $\sim$ 4.5 h$_{75}^{-1}$ kpc.
The velocity field is quite regular.  Fig. 4c shows the rotation curve
rising up to an average maximum velocity of 115 km s$^{-1}$ in 12 arcsec
($\sim$ 4.5 h$_{75}^{-1}$ kpc). This is the largest maximum rotational
velocity we detect for a candidate tidal dwarf galaxy. Interestingly
enough, this is the only tidal dwarf galaxy that seems to be embedded in
a tidal tail and it is the one closest to the parent galaxy. Nonetheless
the very high velocity gradient measured along a very short path and the
optical condensations coincident with the emission-line region motivated
us to list this as a possible candidate. 

{\bf Region 20:} It has an extent of R $\sim$ 3.2 h$_{75}^{-1}$ kpc. As
the profiles of Fig. 3d show, the emission is centrally concentrated. 
The rotation curve presented in Fig. 4d is regular except outside a
radius of 2 h$_{75}^{-1}$ kpc, where we can see a discrepancy between the
approaching and the receding sides. The average maximum velocity 
is 30 km s$^{-1}$. The R image in Fig. 1b shows a single center
and no bridge to the parent galaxy (possibly N7318B).

{\bf Region 21:} It has an extent of R $\sim$ 3.5 h$_{75}^{-1}$ kpc. The
velocity field (Fig. 3e) does not show perturbed isovelocities, but the position
angle changes along the radius. The rotation curve, shown in Fig. 4e, is
derived along a mean position angle of the major axis of $\sim$ 80\degr. The
approaching side of the rotation curve has a normal behavior, with
velocities increasing up to 65 km s$^{-1}$. For the receding side,
due to the position angle change, the velocities are scattered beyond
2 h$_{75}^{-1}$ kpc. 
The R image in Fig. 1b shows two condensations,
the one most to the west being the brightest.

{\bf Region 22:} It shows an extent of R $\sim$ 3 h$_{75}^{-1}$ kpc. The
velocity field is disturbed.  A second component may also be present in
this case, as suggested by the profiles of Fig. 3f. We were not able,
however, to do a satisfactory decomposition of the two components.
The rotation curve, shown in Fig. 4f, presents a strong rise in the first
2", a decrease in velocities after that and then a new increase 
around R $\sim$ 6".  The error bars are very
large for radii beyond 6" indicating a higher scatter in the velocities.
Three condensations are seen in the optical image, the ones most to the
east being the brightest. It is interesting to note that the blueshifted
isovelocities of region 22 point towards the blueshifted isovelocities of
region 21. This suggests the velocity gradients measured are probably
intrinsic to the regions and are not larger scale motions within a
possible faint tidal tail that joins the regions.

{\bf Region 23:} It has an extent of R $\sim$ 2 h$_{75}^{-1}$ kpc. The
velocity field (Fig. 3g) does not show disturbed isovelocities, but the
rotation curve (Fig. 4g) shows that the velocities for the approaching
side increase in the first 2" and then decline while the velocities for
the receding side increase with radii.  The receding and the approaching
sides are therefore not in agreement.

 The rotation curves for four of the galaxies
rise linearly with radius, without reaching a significant plateau. For
two dwarf galaxy candidates, regions 22 and 23, the shape of the curve
is disturbed. For region 6, the shapes of the approaching and receding sides 
are very different.

  The rotation curves presented here
can be compared with those for other tidal dwarf galaxies and blue
compact galaxies. This is done in Section 4.2.

\subsection{H$\alpha$ luminosities and SFR's}

   Table 2 gives H$\alpha$ fluxes for each emitting region, derived
from the calibrated monochromatic map, measured above a threshold of 4.9
$\times$ 10$^{-17}$ erg s$^{-1}$ cm$^{-2}$ arcsec$^{-2}$, 
within apertures of diameters given
in column 9 (defined as the diameter of the aperture of zero intensity).
For region 6 no flux is listed given that the data for this region 
is uncalibrated (see Section 2.1).

   The current star formation rate can be derived from the 
formula SFR(H$_{\alpha}) =
7.5 \times 10^{-8} L_{H_{\alpha}}
(M_{\odot}yr^{-1}L_{\odot}^{-1} $). This formula is valid for 0.1 -- 100
M$_{\odot}$ (Hunter \& Gallagher 1986).  The H$\alpha$ fluxes given in
column 6 of Table 2 were transformed to luminosities for use in the
formula above integrating over the sphere of space, assuming spherical
geometry and a distance to the group of 80 Mpc.

   In contrast to the H$\alpha$ luminosity, the B luminosity is more
sensitive to the older stellar population of a system (Larson \&
Tinsley 1978).  Using the formula proposed by Gallagher \& Hunter (1984)
it is possible to derive the star formation rate averaged for a
period of 10$^9$ years: SFR(L$_B$) = 0.29 $\times$ 10$^{-10} L_B$
($M_{\odot}yr^{-1}L_{\odot}^{-1} $).  The ratio SFR (L$_B$) / SFR
$(L_{H_{\alpha}})$ gives a rough measure of the evolution of the
star-formation rate inside the emission-line region during the last
billion years. The star formation rates are listed in columns 7 and 
8 of Table 2 and are discussed in section 4.3.

\section{Discussion}

%
%

\begin{deluxetable}{lccccccccc}
\small
\tablenum{3}
\tablecolumns{10}
\tablewidth{0pt}
\tablecaption{Photometric and 
kinematic parameters for the tidal dwarf candidates  \label{tbl-3}}
\tablehead{
\colhead{ID} & \colhead{PA} & \colhead{Incl}&  \colhead{L$_B$} & \colhead{$Rh_{75}$} & \colhead{$Dh_{75}$} &
 \colhead{V$_{max}$} &  \colhead{Mass} & \colhead{Mass/L$_B$} & \colhead{M$_{tid}$} \\
\colhead{} & \colhead{$\pm$ (5$^o$)} & \colhead{($^o$)}&  \colhead{(10$^7$ L$_{\odot}$)} & \colhead{(kpc)} & \colhead{(kpc)} & 
\colhead{(km s$^{-1}$)}  & \colhead{(10$^8$ M$_{\odot}$)} 
& \colhead{(M$_{\odot}$/L$_{\odot}$)} & \colhead{(10$^8$ M$_{\odot}$) }}
\startdata
2--5 & 125     & 55 $\pm$ 10 & 6.2 & 6.5  & 23 & 50  & 38  & 6  & 123    \nl
 6     & 130   & 60 $\pm$ 15 & 2.7 & 3.5  & 24 & 50  & 20  & 7   & 12    \nl
 8    &  --5    & 60 $\pm$  5 & 1.9 & 4.5  & 13 & 115 & 138 & 73  & 225   \nl
 20   &  70    & 65 $\pm$ 10 & 0.4 & 3.2  & 23 & 30  & 6.7 & 17   & 15     \nl
 21   &  80    & 60 $\pm$  5 & 1.4 & 3.5  & 27 & 55  & 24  & 17  & 12   \nl
 22     & 75   & 55 $\pm$ 10 & 0.9 & 3.0  & 24 & 25  & 4.3 & 5   & 11   \nl
23      & 40   & 50 $\pm$ 10 & 0.4 & 2.0  & 23 & 23  & 2.4 & 6   & 3.6   \nl

\enddata

\end{deluxetable}

\subsection{Masses, mass-to-light ratios and escape velocities}

The fate of a tidal dwarf galaxy is basically driven by the ratio of its
mass to the virial mass and the ratio of its mass to the so called tidal
mass (Binney and Tremaine 1987, Duc 1995). The virial mass condition tells
us if the tidal dwarf candidate is massive enough to be gravitationally
stable against internal motions.  The tidal mass condition will tell
us if it is massive enough to survive the tidal forces exerted by the
parent galaxy.  In this work we only estimate the tidal mass since for
the virial mass we would need a measurement of the internal velocity
dispersion of the galaxy which we do not have.

 In Table 3 we give, for each dwarf-galaxy
candidate, in (1) the region identification number, (2) the position
angle (3), the inclination (4) the total luminosity of the region, in
the B band, measured inside an aperture of size similar to that within
which the ionized gas was measured, (5) the radius $R$ that corresponds
to the maximum rotational velocity (except for region 22, where $R$
is the radius within which emission was detected), (6) the distance
to the parent galaxy, (7) the maximum rotational velocity (derived
from the rotation curves, shown in Fig. 4), (8) the total galaxy mass,
(9) the mass-to-light ratio and (10) the tidal mass.

  The values given in columns (6), (8) and (10) are based on the
following assumptions:  1) the tidal dwarf galaxy candidate was drawn from
material from galaxy N7318B (or N7319 only for region 6), 2) the dwarf
candidate presents a velocity gradient due to rotation, it is virialized,
and therefore a value for the mass can be derived from the rotation curve,
and 3) galaxies N7319 and N7318B have masses of 1.3 and 1.8 $\times$
10$^{11}$ M$_{\odot}$ respectively (see below).

{\bf Total masses of the tidal dwarf galaxy candidates:}  We estimated
the total mass of the candidate tidal dwarf galaxies from their rotation
curves, derived from the velocity fields (Figs. 3a--3f) using the simple
virial estimator (Lequeux 1983) in which the mass within a radius R of
a rotating disk system is given by M(R)=f $\times$ R $\times$ V(R)$^2$
$\times$ G$^{-1}$. Here V(R) is the rotational velocity at radius R, G
the gravitational constant and f is a constant value between 0.5 and 1.
We chose a value of 1 for simplicity.  

M(R) is a lower limit on the estimate of the mass, since most likely
these regions are not pure rotators.  Random  motions of the gas could
provide significant dynamical support for the tidal dwarf galaxies as
indicated by their line widths.  As an approximation, if we assume that
the gas is in equilibrium, and that the gas velocity ellipsoid is
isotropic, a very rough estimate of the contribution from the pressure
support could be made using an ``asymmetric drift correction''
following Oort (1965). For gaseous objects in formation, such as those
present in the Stephan's quintet, and under the above assumptions, a
general rule may be:  1) for rotation curves with amplitudes V$_{max}$
$\sim$ 25 km s$^{-1}$, i.e.  for the most internal regions of the self
gravitating structures or for the least massive objects, the
contributions from the dispersion and the rotation components may be
comparable; 2) for V$_{max}$ $\sim$ 50 km s$^{-1}$, the contribution of
the dispersion velocity component may be approximately half of that for
the rotational velocity component and 3) for V$_{max}$ $>$ 50 km
s$^{-1}$, i.e. at larger distances from the center of the objects and/or
for more massive structures, the dispersion velocity may be negligible
at first approximation.

{\bf M/L of the tidal dwarf galaxy candidates:}  The mass-to-light
ratios of the candidate dwarf galaxies were obtained by dividing the
values in column 8 by those in column 4 of Table 3.  The median value
for the mass-to-light ratios of the tidal dwarf galaxy candidates is
7 $M_\odot/L_\odot$, with a large scatter, with values as high as 74
$M_\odot/L_\odot$ for region 8. The mass-to-light ratio of region 8
reflects its steep velocity gradient of $\sim$ 26 h$_{75}$ km s$^{-1}$
kpc$^{-1}$, the highest observed among the tidal dwarf candidates observed
(see a discussion of this point in section 4.4).

{\bf Total masses of the progenitor galaxies:} In order to determine
the tidal masses, an estimate of the total
mass of the progenitor galaxies is necessary.  As we were not able to
measure the rotation curves of the parent galaxies N7319 and N7318B due
to the lack of ionized gas in their disks, we estimated their masses
using the Tully-Fisher relation
(M$_B$ = --5.85 log (2 V$_{max}$) --
5.61, Pierini (1999),
the factor 2 within the log was included to match our
definition of V$_{max}$ to theirs). 
 For absolute magnitudes of --21.4 and --21.3 and R$_{25}$ of 10.4 and 
13.2 h$_{75}$ kpc 
for N7319 and N7318B respectively we obtain maximum velocities of 249
and 240 km s$^{-1}$.  Using the virial estimator we then obtain masses of
1.3 and 1.8 $\times$ 10$^{11}$ M$\odot$.

{\bf Tidal masses:}  The tidal masses listed in column 10 of Table 3
were derived from Binney and Tremaine (1987), M$_{tid}$ = 3M(R/D)$^3$,
where $M$ is the mass of the parent galaxy, $R$ the radius of the tidal
dwarf and $D$ the distance to the parent galaxy.  We assumed that the
masses of the parent galaxies NGC 7318B (for regions 2-3-4-5, 8, 20,
21, 22 and 23) and NGC 7319 (for region 6) are 1.3 and 1.8 $\times$ 10$^{11}$
M$_{\odot}$ respectively, as calculated above.  The results in Table
3 show that the dwarf masses determined are in two cases larger than
the tidal masses (for regions 6 and 21), suggesting that the regions
considered may survive tidal forces. This is not the case, however,
for the complex of regions 2--5 and for regions 8, 20 and 22, for
which the tidal mass is larger than the total mass estimated.

{\bf Escape velocities of the candidates:} 
We can also estimate the escape velocity of the tidal dwarf candidates
and compare this velocity to the systemic velocity difference between
the dwarf candidate and the parent galaxy.  We have found that
the velocity difference between region 21 and N7318B is of the order
of the escape velocity calculated at the distance of the region (27
h$^{-1}$ kpc), indicating that it may be able to escape from the parent
galaxy potential. This is not the case for regions 2--5 and 6 (for the
latter N7319 is the parent galaxy).  For regions 8, 20, 22 and 23 we
do not have an unique systemic velocity determination. Nevertheless
we can give some estimate based on the possible systemic velocities
(see Plana et al. 1999).  For these four regions (8, 20, 22 and 23),
the minimum possible velocities are $\sim$ 5935, 5855, 5975 and 5935
km s$^{-1}$, respectively (Plana et al. 1999, their Table 1).  If we
assume that NGC 7318B (with velocity 5774 km s$^{-1}$) is the parent
galaxy, then the differences between the velocity of the parent galaxy
and that for each of the tidal dwarf candidates are 161, 81, 201 and
161 km s$^{-1}$ (or larger) for regions 8, 20, 22 and 23 respectively.
On the other hand, the escape velocities are 344, 259, 253 and 259 km
s$^{-1}$ respectively.  Therefore, if the minimum possible velocities
are the true radial velocities, only region 22 of the four may be able
to escape the potential of the parent galaxy, since for this region the
escape velocity is similar to the velocity difference.  We then conclude
that only two out of the seven candidates may have velocities such that
they can escape the gravitational field of the parent galaxy.

\subsection{Comparison of the velocity fields}

  Just a few kinematic studies are available on tidal dwarf galaxies
and dwarf galaxies in general.  Duc and Mirabel (1998) showed velocity
curves of dwarf galaxies around NGC 5291 measured from HI kinematics.
The curves show in general stronger gradients than those measured for
the galaxies in Stephan's quintet.  However, the diameter of the objects,
as computed from the optical images, are similar.

Another set of velocity curves and rotation curves for dwarf galaxies
comes from \"Ostlin et al. (1999) and \"Ostlin et al. (2000).  Their
sample is composed of luminous blue compact galaxies (hereafter BCG's).
The radial extent of the rotation curves are comparable to those for
the candidate tidal dwarf galaxies in Stephan's quintet and the velocity
amplitudes also seem to be similar.  The shapes of the rotation curves,
however, are very different.  While in our sample the rotation curves show
in general a typical shape for solid body rotation,
in \"Ostlin et al. (1999)
the rotation curves of the BCG's have flat plateaus.

In \"Ostlin et al. (1999, 2000), the velocity fields are irregular and
often contain secondary dynamical components although they in most cases
display overall rotation.  By comparison of the stellar masses (by means
of multicolor images and spectral evolutionary synthesis' analysis) to
the total masses of the galaxies (derived from the rotation curves), they
showed that at least half of the galaxies cannot be supported by rotation.
They found that the morphologies and dynamics of the BCG's suggest that
the starburst activity in these galaxies are most likely triggered by
mergers of dwarf galaxies and/or massive gas clouds.  In the case of
the candidate tidal dwarf galaxies, it is possible that they are also
partially supported by random motions as it is the case for the BCG's.
Moreover, the starburst activity in the tidal dwarf galaxies could also
be nurtured by merging of gas clouds during the ongoing tidal process.

   Hunter et al. (2000) have recently made a compilation of possible
dwarf galaxies in many different environments which could have tidal
origin. A comparison of the maximum rotation velocities as a function
of the B absolute magnitudes of the tidal dwarf galaxy candidates in
Stephan's quintet and galaxies in Hunter's sample is shown in Fig. 5. As
can be seen, the velocities and magnitudes of the H92's tidal dwarf
candidates are within the ranges of parameters of other possible
candidates.

\subsection {Comparison of the SFR for various classes of emission-line
objects}
  
    There are several classes of emission-line objects which have
similarities in some respects with the objects we observed in Stephan's
quintet.

  A comparison of the star formation rates (SFR) for five samples of
dwarf galaxies, one sample of HII regions in nearby spiral galaxies
and our sample is shown in Fig. 6. We identified each of the galaxies
in our sample with its corresponding ID number.  Included in the
comparison are (1) a sample of irregular galaxies from the Local Group
(Mateo 1998), (2) a sample of isolated irregular galaxies of low
luminosity and central surface densities, (3) a sample of blue compact
galaxies studied by Sage et al.  (1992) with blue luminosities similar
to the systems observed in Stephan's quintet, (4) another sample of
blue compact galaxies studied by \"Ostlin et al. (1999), (5) a sample
of tidal dwarf galaxy candidates around N5291 studied by Duc and
Mirabel (1998), and (6) a sample of HII regions in the spiral galaxies
NGC 1365, NGC 1566, NGC 2366, NGC 2903, NGC 2997, NGC 3351, NGC 4303,
NGC 4449, NGC 5253 studied by Mayya (1994, their table 1, only objects with
data quality 1 or 2 were included).

    Fig. 6 shows that the Local Group and isolated irregular galaxies have
SFR($H\alpha$)'s that are on average lower than those for all the other
samples. HII regions in nearby galaxies and blue compact galaxies have
values for the SFR($H\alpha$) in between those observed for isolated
galaxies and those for tidal dwarf galaxies and giant HII regions of
Stephan's quintet.  While in the sample of Sage et al. (1992) the blue
compact galaxies have a comparable range of luminosities to the galaxies
in the other samples, \"Ostlin et al.'s sample is dominated by blue
compact galaxies with high B-luminosities. Still, the SFR($H\alpha$)
for these two samples agree in the region of overlap.  The faintest blue
compact galaxies seem to overlap with the location of the figure occupied
by the HII regions in nearby spiral galaxies, although the latter extend
to much lower luminosities.  The tidal dwarf galaxies studied by Duc \&
Mirabel (1998) have average SFR($H\alpha$)'s that are comparable to the
objects studied in this paper. However, Duc \& Mirabel's measurements
were made through long slit observations and the values may be a lower
limit on the H$\alpha$ fluxes and therefore to the SFR($H\alpha$)'s
derived for these galaxies.

  The SFR($H\alpha$)'s for the BCG's may be affected by dust, which
would artificially lower the rates derived for these objects. Therefore,
in reality, the SFR($H\alpha$)'s for tidal dwarf galaxies and BCG's may be
similar although these galaxies are fundamentally two distinct classes
(for instance they have completely different metallicities, with the
BCG's being metal poor objects and the tidal dwarf galaxies having the
high metallicities of their parent galaxies). One should also be aware
that there exists a very large range of SFR($H\alpha$)'s for both the
BCG's and the tidal dwarf galaxies. Only a crude comparison between the
SFR($H\alpha$)'s of the two populations is then possible.

\subsection{Considerations about streaming motions in tidal tails}

 Region 8 as well as the complex of regions 2--5 seem to be a part of
a tidal tail. In addition, region 6 may be at the edge of a tidal tail,
if we assume that it is associated with N7318A and not with N7319
(see section 4.7).  One obvious concern is then that the observed
velocity gradients for these regions could be a result of streaming
motion in the tail within which the region is embedded.

  In the case of region 8, if we assume that it is
attached to N7318B (at a velocity of $\sim$ 5700 km s$^{-1}$) and it has a
velocity of $\sim$ 5900 km s$^{-1}$ (J. Sulentic, private communication),
the gradient of the motion in the tidal tail (lower velocities in the
south and higher velocities in the north) has opposite sense to what
is observed in the region itself (the red side of the region is to the
south, see Fig. 3c). This represents clear evidence that the strong
velocity gradient observed for region 8 cannot be due to streaming
motions within the tidal tail.  In addition, we note that there is a
large misalignment of at least 30 degrees between the PA of the major
axis of region 8 and that of the tidal tail within which it seems to
be embedded (see Fig. 1a). This again indicates that region 8 cannot be
just a part of a tidal tail and it is, instead, an object with
its own rotation pattern.

   In the case of the complex of regions 2--5,
we find that the velocity gradient we observed for it (these regions are
located at the edge of a tidal tail that emanates from the northern arm
of N7318B) could not be strongly affected by streaming motions for the
following reason.  For this complex we have information on the HI velocity
field kindly provided by L. Verdes-Montenegro (private communication).
The motion observed in HI encompasses an area that goes from the base
of the northern tail of N7318B to our region 1 and it has a very similar
mean radial velocity ($\sim$ 6000 km s$^{-1}$) to that of the H$\alpha$
emission in complex 2--5.  We compared the H$\alpha$ and the HI velocity
fields for regions 2--5 and we found that while the H$\alpha$ gradient
reaches 50 km s$^{-1}$ over a radius of 13 arcsec in extent, the HI
gradient reaches only about 15 km s$^{-1}$ over the same area. Assuming
that the HI velocity gradient gives us an upper limit on the large-scale
motion of the tidal tail (upper limit because some internal motion is
also expected to be present in the HI component) the difference between
the two gradients (35  km s$^{-1}$) is due to internal motions within
the complex 2--5.  With this correction the new mass-to-light ratio of
the complex of regions 2--5 would be 3 $M_\odot/L_\odot$.

   In a scenario where region 6 were at the edge of the tidal tail that
emanates from N7318A, there would be no concern about its velocity
gradient being due to streaming motion since the center of the parent
galaxy and region 6 have almost exactly the same velocity.

   We conclude that streaming motions within the tails
may not significantly change the observed velocity gradients measured
in this study.

\subsection {Other dynamic considerations}

  Dubinski et al. (1999) find
that a good criterion for making tidal tails during a collision appears
to be that the ratio of the escape velocity (V$_e$) over the circular
velocity (V$_{max}$, in our notation) of the galaxy, measured at a
radius of 2R$_d$, must be lower than 2.5 (R$_d$ is the scale length of
the galaxy).  According to Dubinski et al. (1999) this condition is
valid for a wide range of models including those with disk-dominated
and halo-dominated rotation curves. They also find that galaxies with
declining rotation curves are the best candidates to produce tidal tails
during collisions.  We do not have information on the rotation curves
of the galaxies (due to their lack of ionized gas) but we can still
check the velocity ratio condition V$_{esc}$/V$_{max}$ $<$ 2.5 assuming
that the galaxies follow the Tully-Fisher relation (see Section 4.1).
The masses inside 2R$_d$ are 1.1 and 0.8 $\times$ 10$^{11}$ for N7318B and
N7319 respectively (if we assume that the scale length of the galaxies
R$_d$ $\sim$ R$_{25}$/3.2). The ratio  V$_{esc}$/V$_{max}$ is, then,
well below 2.5 for both galaxies, indicating that they may develop tidal
tails during collisions. This is a comfortable confirmation since both
galaxies obviously have developed one or more tidal tails.

 It is interesting to note that the inclination of the seven tidal dwarf
candidates lie in between the values of 50\degr and 65\degr, which are
very similar to the inclinations of the progenitor galaxies of 59\degr and
66\degr (from the Lyon-Meudon Extragalactic Database, LEDA), for N7319 and
N7318B respectively.  If the candidate tidal dwarf galaxies are formed
in the plane of the disks of the parent galaxies, they probably could
keep the angular momentum of their progenitor disks.  This inclination
concordance would then be expected by the conservation of angular momentum
of the system.  However, the inclination found for the candidates could
be biased by the fact that standard methods for deriving inclinations
favor high values of inclination (low inclination values are harder
to be determined). In addition, since we do not have information on
which side of the galaxy is pointing to us, the two inclined disks with
60\degr inclinations could actually be separated by 120\degr!

\subsection{Summary of the properties of the tidal dwarf galaxy candidates}

%
%

\begin{deluxetable}{lccccccc}
\small
\tablenum{4}
\tablecolumns{8}
\tablewidth{0pt}
\tablecaption{Summary of the properties of the tidal dwarf galaxies  \label{tbl-4}}
\tablehead{
\colhead{ID} & \colhead{Blue} & \colhead{SFR(H$\alpha$)/SFR(L$_B$)}& \colhead{Velocity Gradient}& \colhead{M$_{tidal}$} & \colhead{V$_{esc}$} & 
\colhead{Location} & associated w/ \\
\colhead{} & \colhead{Color} & \colhead{condition}& \colhead{condition}&\colhead{} & \colhead{condition}&\colhead{}& \colhead{HI cloud?}}
\startdata
2--5 & + & + & + & -- & -- & End of tail & +\nl
6       & -- & + & + & + & -- & End of tail & + \nl
8       & + & + & + & -- & -- & Inside a tail & -- \nl
20      & + & + & + & -- & -- & Isolated & -- \nl
21      & + & + & + & + & + & Isolated & -- \nl
22      & + & + & + & -- & + & Isolated & -- \nl
23      & + & + & + & -- & -- & Isolated & -- \nl

\enddata

\end{deluxetable}

 Table 4 summarizes the characteristics of each of the seven tidal
dwarf candidates of Stephan's quintet. A ``plus'' in an entry means
(1) In column two if the color is bluer than B--R = 0.8, (2) in column
3 if the SFR(H$\alpha$)/SFR(L$_B$) (see section 3.4) is typical of a
tidal dwarf galaxy, (3) in column 4, if the galaxy presents a velocity
gradient larger than  8.0 h$_{75}$
km s$^{-1}$ kpc$^{-1}$ (this will be true for all of the objects since
they were classified as tidal dwarf candidates based on the presence of
the velocity gradients), (4) in column 5 if the mass of the galaxy as
measured from the internal kinematics of the ionized gas is larger than
the tidal mass obtained in section 4.1, (5) in column 6 if the escape
velocity of the parent galaxy is lower than the velocity difference
between the systemic velocity of the tidal dwarf candidate and that of
the parent galaxy (assuming that the parent galaxy is N7318B for all
regions except for region 6, for which the parent galaxy is assumed to
be N7319). The last two columns of Table 4 mark if the candidate belongs
to a tidal tail and if it is associated with an HI cloud.

   Regions 6 and 21 seem to be the ones with the highest chance of
surviving as single entities. This is specially true for region 6 where
CO emission at the redshift and location of the galaxy has been recently
detected (Gao \& Xu 2000) indicating a molecular gas mass of a few times
10$^8$ M$_{\odot}$.  Other less massive candidates like regions 20,
22 and 23 may not be massive enough to become independent entities.
However, we should keep in mind that the values for the mass were
determined under the assumption the regions are pure planar rotators
and hence these are lower limits.  In fact, random  motions of the gas
and stars may provide significant dynamical support for the tidal dwarf
galaxies as indicated by their line widths.  Larger values for the mass
of the candidates (obtained if the internal dispersion velocity is also
taken into account) would result in an increase of the number of
objects that could survive.

  Although the complex of regions 2--5 may not survive as a
single entity (the tidal mass is very large mainly because of the large
radius of the complex), it may be split into several smaller candidates
which could then survive the tidal forces exerted by N7318B.

\subsection{Other probable tidal dwarf galaxy candidates}

    The HI map of Stephan's quintet presented by Williams et al. (1999,
their Fig. 5) shows that the HI component at velocity 6600--6700 km s$^{-1}$ is
distributed along a fragmented ring around the galaxy N7319, possibly
formed after a tidal interaction with N7320C.  Such HI rings have
previously been observed in other interacting systems like e.g. N5291
(Malphrus et al. 1997).  The northwestern side of the HI ring overlaps
with the dwarf galaxy candidate identified as Region 6.  It is possible
that there are other tidal dwarf galaxy candidates along what would be
the continuation of the ring, to the north, beyond region 6, north of
N7319, where HI is not measured by Williams et al. (1999).  

  Comparing the eastern side of the HI ring with an optical image  
of the quintet we notice that the end of the tidal tail of N7319 that
points to N7320C (in the southeastern side of the group) has a number
of optical condensations (Hunsberger et al. 1996, Gallagher et al. 2000)
that overlaps with maxima in the HI distribution. These regions probably
have similar velocities to that of the HI gas (6700 km s$^{-1}$) and
are most likely tidal dwarf candidates formed during a past encounter
between N7319 and N7320C.  Other faint condensations can be seen in this
region, between N7319 and N7320C. The velocities of these objects have
to be measured in order to confirm their nature.

  What is very intriguing at first is the exact superposition of the HI
component at 6700 km s$^{-1}$ with the galaxy NGC 7320, at 800 km
s$^{-1}$.  This superposition may be understood if the following is
considered.  Moles et al.  (1998) noted that while in a narrow-band
image (e.g. H$\alpha$) the galaxy NGC 7320 appeared symmetric, in a
broad-band image it appeared very asymmetric, showing an extension to
its northwest.  In fact, what may appear to be the northwestern extension
of the nearby galaxy N7320 in broad-band images may be in reality a
chain of smaller background objects (perhaps other tidal dwarf galaxy
candidates?) superimposed on the line of sight, but with velocities that
coincide with that of the HI component at 6700 km s$^{-1}$. Other regions
may have velocities of $\sim$ 5700 km s${^-1}$ (possibly associated with
N7318B).  Support for this scenario comes from the mix of velocities
measured in this region by Moles et al. (1998).  Again, more velocity
measurements of the condensations present in this region are needed.

  Other tidal dwarf galaxies may be associated with N7318A. Although
Moles et al. (1998) did not consider this galaxy in the interacting
history of the quintet, it must have been part of collisions given the
peculiar pattern of dust seen in the HST image (Charlton et al. 2000)
and due to the several linear features that seem to emanate from the
galaxy. These were marked with arrows in Fig. 1c. A few of these could
be projected material from N7318A into the intragroup medium, in the
same way it has been observed for Arp 105 by Duc (1995). Alternatively
it could be material in the process of re-accretion onto the parent
galaxy after an interaction. There seems to be two different types of
linear features, a few that are long and filamentary, with very low
surface brightnesses and others that are small and have high surface
brightnesses.  Perhaps also region 6 may be born out of gas from N7318A
and not of N7319 as it has been assumed throughout this paper. This
would be a specially appealing scenario if we could confirm that there
is material in between galaxy N7318A and region 6 at a common velocity
(of $\sim$ 6700 km s$^{-1}$). However, it is more probable (see Fig. 2 of Xu et
al. 1999) that the apparent optical connection between regions 8 and 4
(seen as a filament in the HST image) is at a lower velocity of 5900--6000
km s$^{-1}$.  Also important would be to confirm if the features marked
in Fig. 1c coincide with the velocity of N7318A.

  We would also like to point out that the emission detected
in regions 7 and 9 and in the features in between these two regions
might be associated with the higher velocity component of the group
at 6700 km s$^{-1}$ instead of being related to the lower velocity
component (as it has been assumed in this paper).  This is supported
by the H$\alpha$--[NII] map presented by Xu et al. (1999, their Fig. 2),
which shows clearly that these regions are dominated by emission from the
high-velocity component. Further indirect evidence is their low H$\alpha$
content despite their high optical luminosities (see Fig. 1a and 1c),
which suggests that the emission from this area may perhaps fall at
the very edge of the CFHT filter, where the transmission is minimum.
These regions would not have been detected in the lower S/N data taken
with the SAO 6m, where only very bright sources like region 6
were detected.  As can be seen from Fig. 1c the {\it optical} luminosities
of the sources in the area in between regions 7 and 9 are comparable
to those of other regions identified by us.  This only illustrates the
incredible mixture of matter at two different velocities in this group.

\subsection{Properties of region 6 in a scenario where N7318A is the parent galaxy}

    We assumed in the previous sections that region 6 is formed of remnant
gas from N7319 (Moles et al. 1997). We favored this scenario based on
the evidence from the HI observations (Fig. 5 from Williams et al. 1999)
which showed a structure similar to a ring (an HI incomplete ring) around
galaxy N7319. The HI gas associated with region 6 seems to be part of this
large ring, in which case region 6 would be formed from gas originally
in N7319.  On the other hand, there is also the possibility that region
6 was formed from gas spelled from N7318A. Support for this is given by
Fig. 1c, where linear features apparently associated with N7318A are seen.
In this scenario region 6 might be located at the tip of a tidal tail
that may originate at the northern arm of N7318A. We need new velocity
measurements of the region to decide between the two scenarios.

  In the following we give the parameters for region 6 in the case where
it is physically associated with N7318A: (1) it would be at a distance of
25 h$_{75}^{-1}$ kpc to the parent galaxy; 
(2) it would have a tidal mass of 6 $\times$ 10$^{8}$ M$_{\odot}$,
smaller than its inferred mass; (3) the difference between its radial
velocity and that of the parent galaxy is still lower than the escape
velocity of the parent galaxy; (4) its inclination (60\degr $\pm$ 15\degr)
would still be similar to (within the errors) to that measured for the
parent galaxy (49\degr, value taken from the Lyon-Meudon Extragalactic
Database, LEDA).

\section{Summary}

We presented B and R photometry and a Fabry-Perot H$\alpha$ 
map of 23  emission-line
regions in Stephan's quintet. All but one region 
may be associated with N7318B (region 6 may be attached either to N7319 
or to N7318A).  

Our main results are:

\begin{itemize}

\item We find that seven of the regions have velocity gradients greater
than 8 h$_{75}$ km s$^{-1}$ kpc$^{-1}$. 
We classify those as tidal dwarf galaxy candidates.

\item   Two tidal dwarf candidates may be located at the edge of a tidal tail,
one within a tail and for four others there is no obvious stellar/gaseous
bridge between them and the parent galaxy.

\item  Two of the candidates are
associated with HI clouds, one of which is, in addition, associated with 
a CO cloud.

\item The tidal dwarf candidates have low continuum fluxes and high H$\alpha$
luminosity densities of F(H$\alpha$) = 1 -- 60 $\times$ 10$^{-14}$ erg
s$^{-1}$ cm$^{-2}$,  magnitudes of M$_B =$ --16.1 to --12.6, sizes of
typically $\sim$ 3.5 h$_{75}^{-1}$ kpc, colors $B-R$ between  0.3 and
1.3, gas velocity gradients of $\sim$ 8 -- 26 h$_{75}$ km s$^{-1}$ kpc$^{-1}$,
SFR(H$_\alpha$)/SFR($L_B$) between $\sim$ 1100 and 6000, masses of $\sim$ 2
$\times$ 10$^8$ to 10$^{10}$ M$_\odot$ and a median mass-to-light ratio of
7  M$_\odot$/L$_\odot$.

\item The lower limits of the masses of the candidate tidal dwarf
galaxies determined from their rotation curves (assuming that they are
pure rotators) are in two cases larger than their tidal masses.

\item Two out of the seven candidates may have velocities such that they can
escape the gravitational field of the parent galaxy.

\item Possible streaming motions within the tails may not have
significantly affected the observed velocity gradients of those tidal
dwarf candidates that are located within tails.

\item The dynamical criterion for formation of tidal tails during a
collision V$_{esc}$/V$_{max}$ $<$ 2.5 is 
followed by galaxies NGC 7319 and NGC 7318B. 

\end{itemize}

 A few of the tidal dwarf galaxies we identified
in this study may survive as single entities, leading to the formation
of new group members of the Stephan's quintet.

\acknowledgements

We would like to thank Dr. Jacques Boulesteix and the SAO observatory
staff (Drs V. Afanasiev, A. Burenkov, S. Dodonov, V. Vlasiuk and
S. Drabek) for their help during the observations. We would also like
to thank Drs. S. Hunsberger, M. Marcelin and J. Sulentic for insightful
discussions and Lourdes Verdes-Montenegro for sending us the HI map of
the quintet.  CMdO thanks the Marseille Observatory for their hospitality
and funding of a trip in July 1999.  PA thanks the Brazilian funding
agency FAPESP for the financial support for two visits to the Instituto
Astronomico e Geofisico, project numbers 99/05514-3 and 99/03744-1.
HP acknowledges the financial support of FAPESP through a pos-doc
position, project number 96/06722-0, and the Mexican CONACYT under
project 32303-E.

%
%

\begin{figure*}
\figurenum{1a}

\plotfiddle{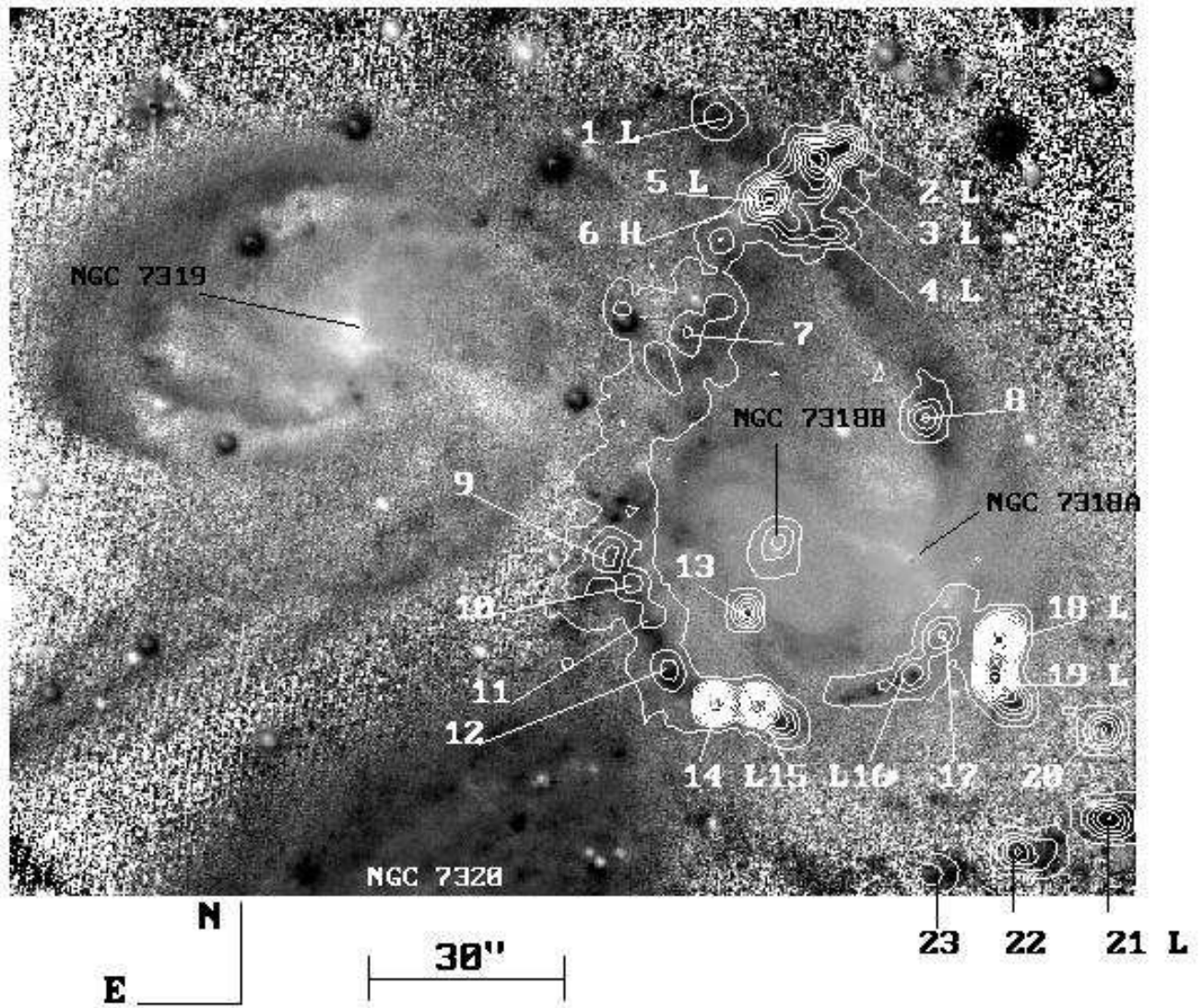}{14cm}{0}{90}{90}{-260}{-100}

\caption{(a) This figure presents the B--R color
image. Dark is blue and white is red. The dimension of the image is
3 $\times$ 2.4 arcmin. North is up and east is to the left. The H$\alpha$
monochromatic emission contours are shown and the 23 emitting regions
detected by Plana et al. 1999 are marked. An ``L'' (for low) marks regions
known to have velocities in the range 5700--6000 km s$^{-1}$ while an
``H'' (for high) marks region 6, the only region for which the velocity
is confirmed to be $\sim$ 6700 km s$^{-1}$. The monochromatic contours
are flux calibrated. The lowest level is 4.2 $\times$ 10$^{-17}$ erg
s$^{-1}$ cm$^{-2}$ arcsec$^{-2}$ and the step is 1.7 $\times$ 10$^{-17}$
erg s$^{-1}$ cm$^{-2}$ arcsec$^{-2}$. (b) This figure presents the R
image of the group after sky subtraction. Optical blobs corresponding to
H$\alpha$ emitting regions are marked. (c) 
The final combined B image of the quintet.
Linear features which could possibly be associated with N7318A are 
marked with arrows (see text).}
\end{figure*}
\newpage

%
%

\begin{figure*}
\figurenum{1b}

\plotfiddle{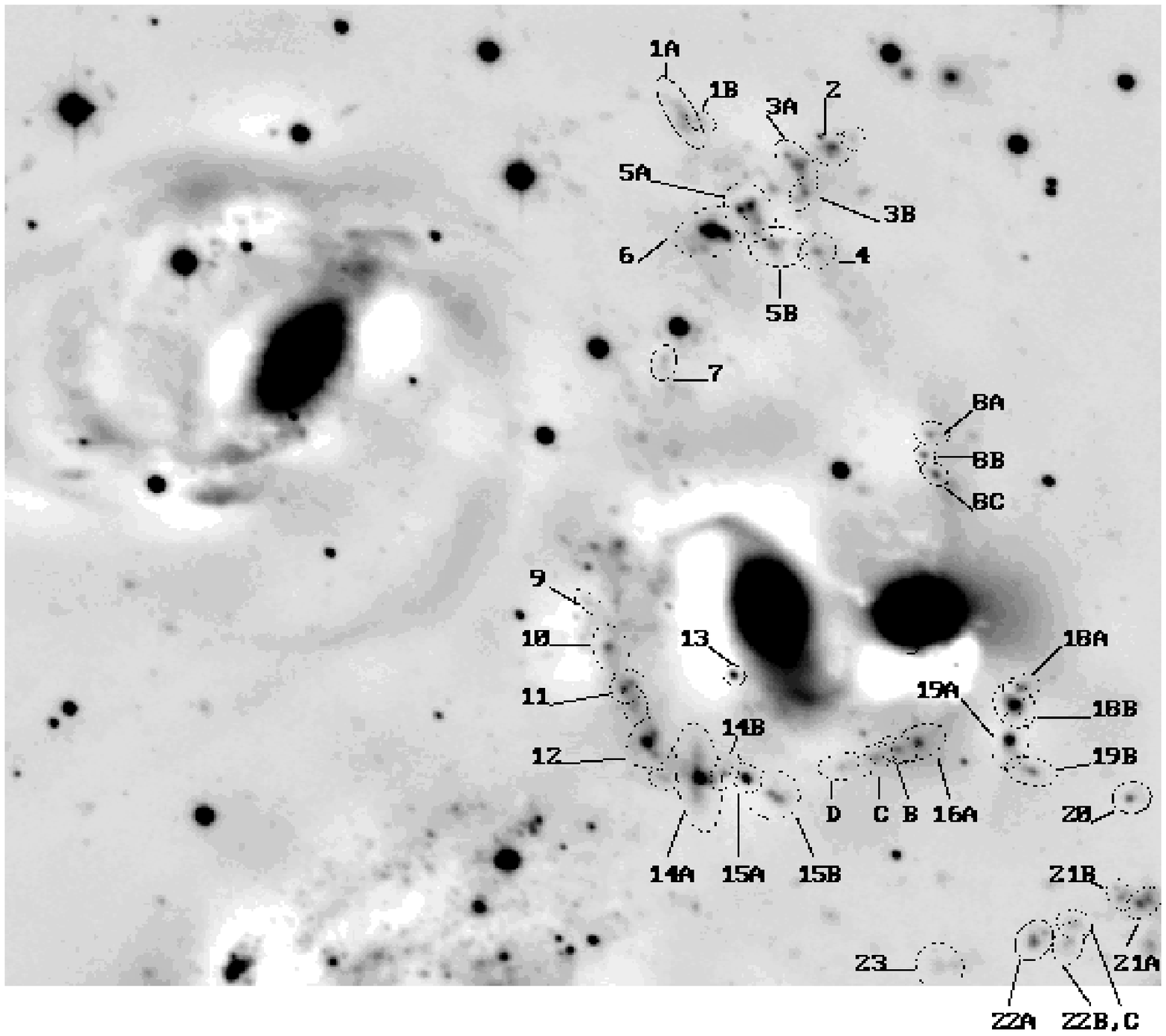}{14cm}{0}{90}{90}{-250}{-100}

\caption{}
\end{figure*}
\newpage

%
%

\begin{figure*}
\figurenum{1c}

\plotfiddle{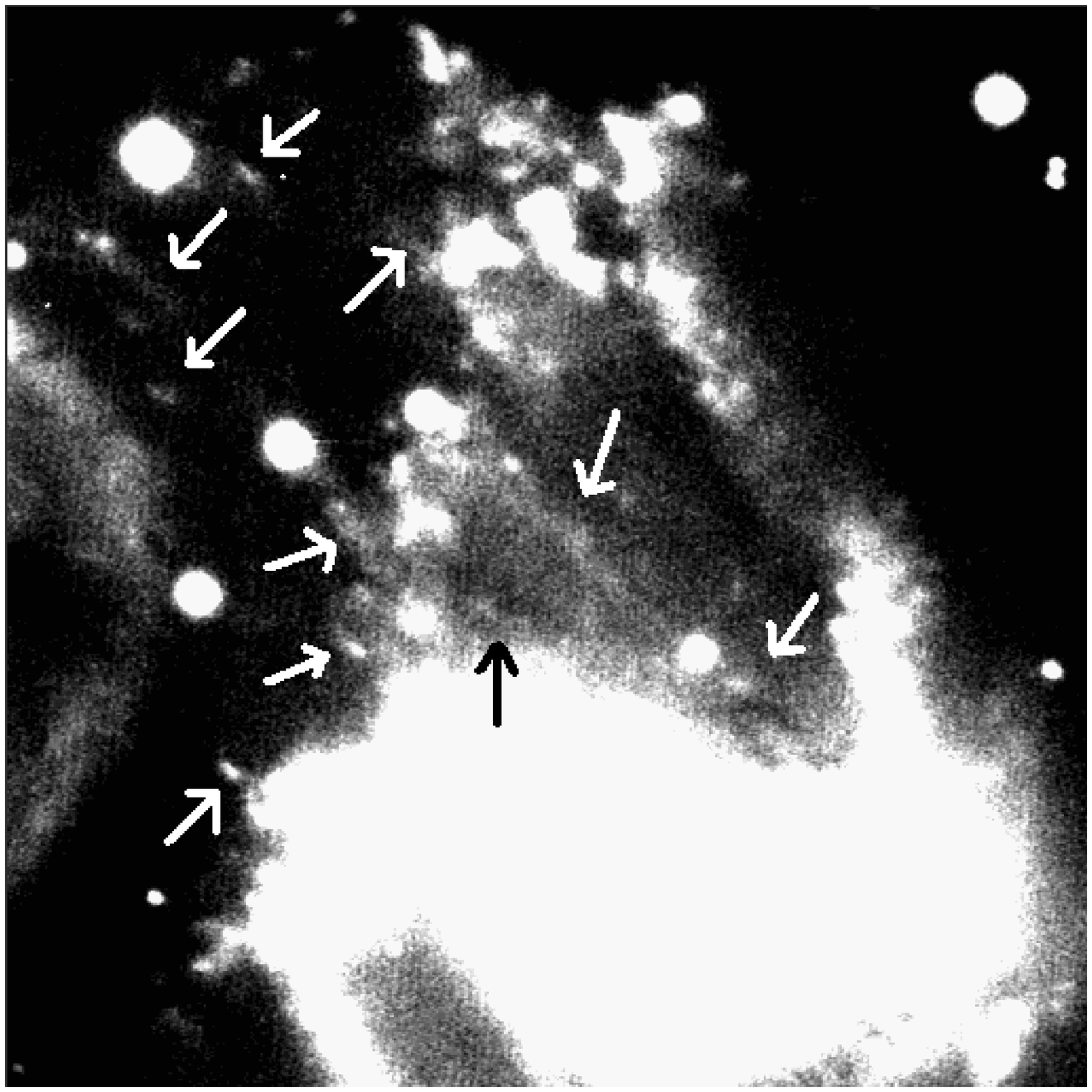}{14cm}{0}{90}{90}{-260}{-100}

\caption{}
\end{figure*}
\newpage

%
%

\begin{figure*}
\figurenum{2}

\plotfiddle{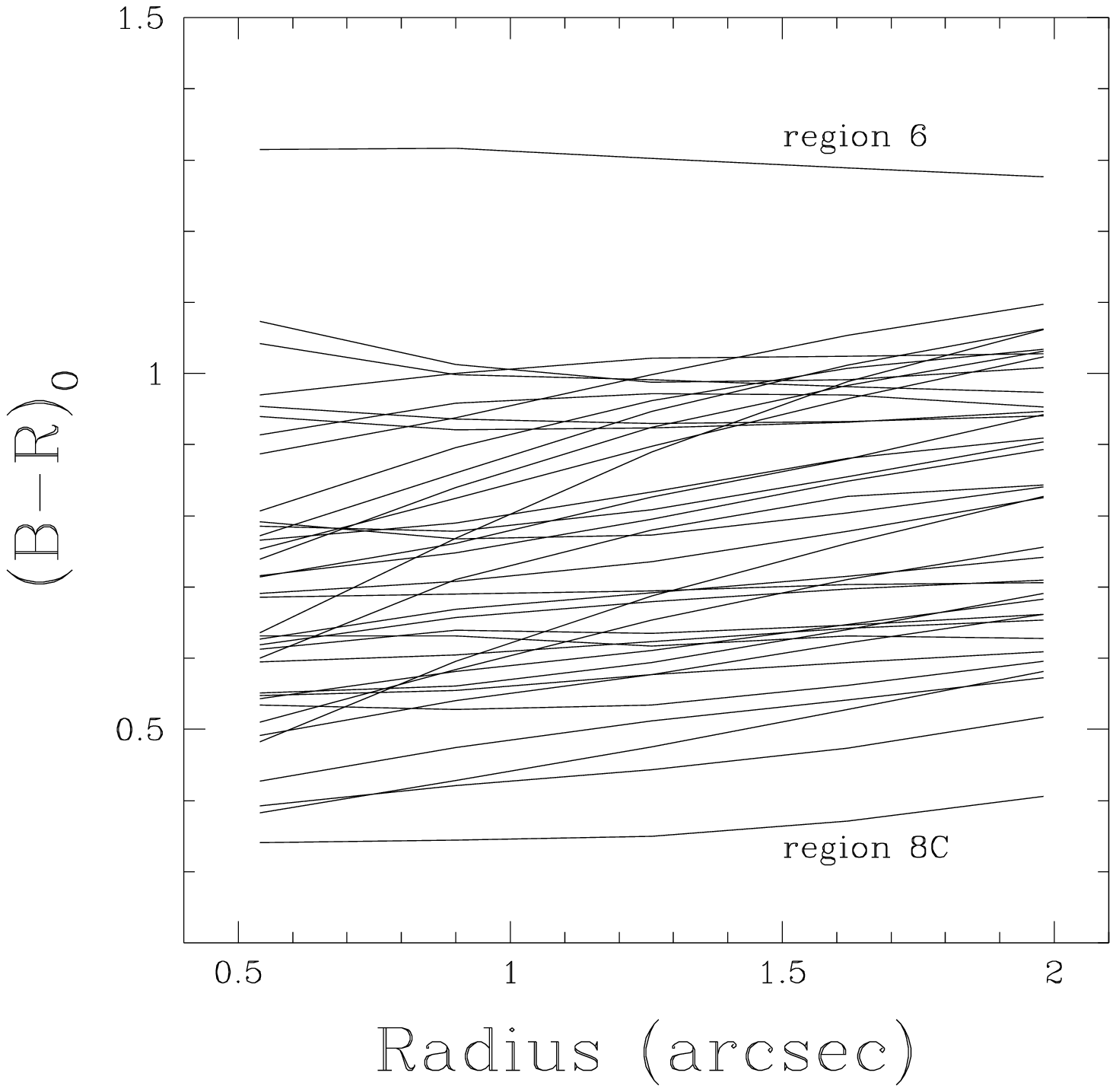}{14cm}{0}{90}{90}{-300}{-200}

\caption{The corrected
B--R color gradient for each emission-line blob identified 
in Fig. 1b (see text). }
\end{figure*}
\newpage

%
%

\begin{figure*}
\figurenum{3ab}

\plotfiddle{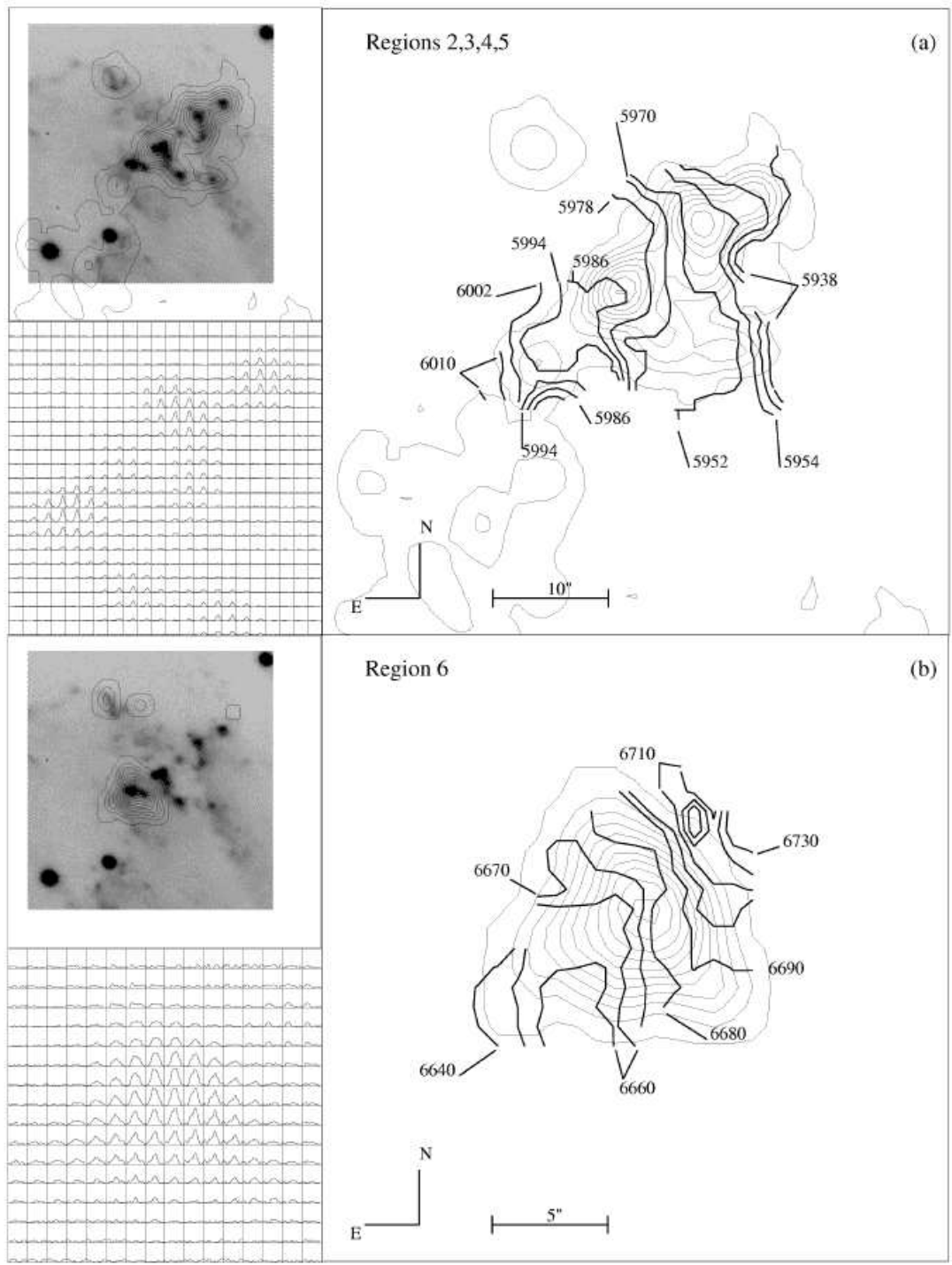}{14cm}{0}{77}{77}{-230}{-35}

\caption{From (a) to (g) each figure presents the
following. Upper left: B band
image superimposed onto the monochromatic map; the isocontours are the 
same as those presented in Fig. 1a; lower left: profiles from the 
Fabry-Perot
data cube -- the pixel size is 0.86;
right panel: the 
velocity field superimposed onto the monochromatic image.
The field size 
for the right panel is the same as that for the upper left panel.}
\end{figure*}
\newpage

%
%

\begin{figure*}
\figurenum{3cd}

\plotfiddle{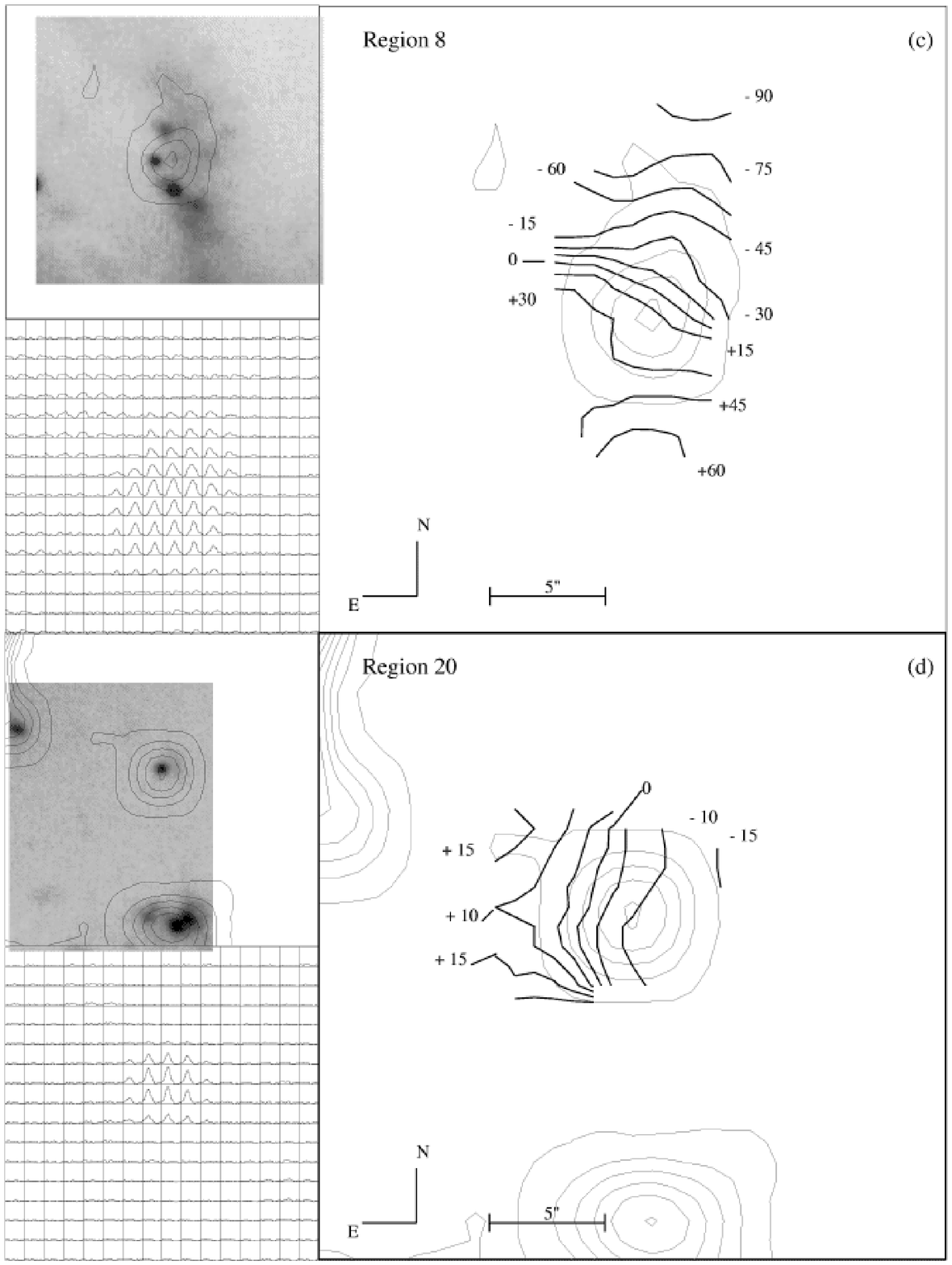}{14cm}{0}{77}{77}{-230}{-35}
\caption{}
\end{figure*}
\newpage

%
%

\begin{figure*}
\figurenum{3ef}

\plotfiddle{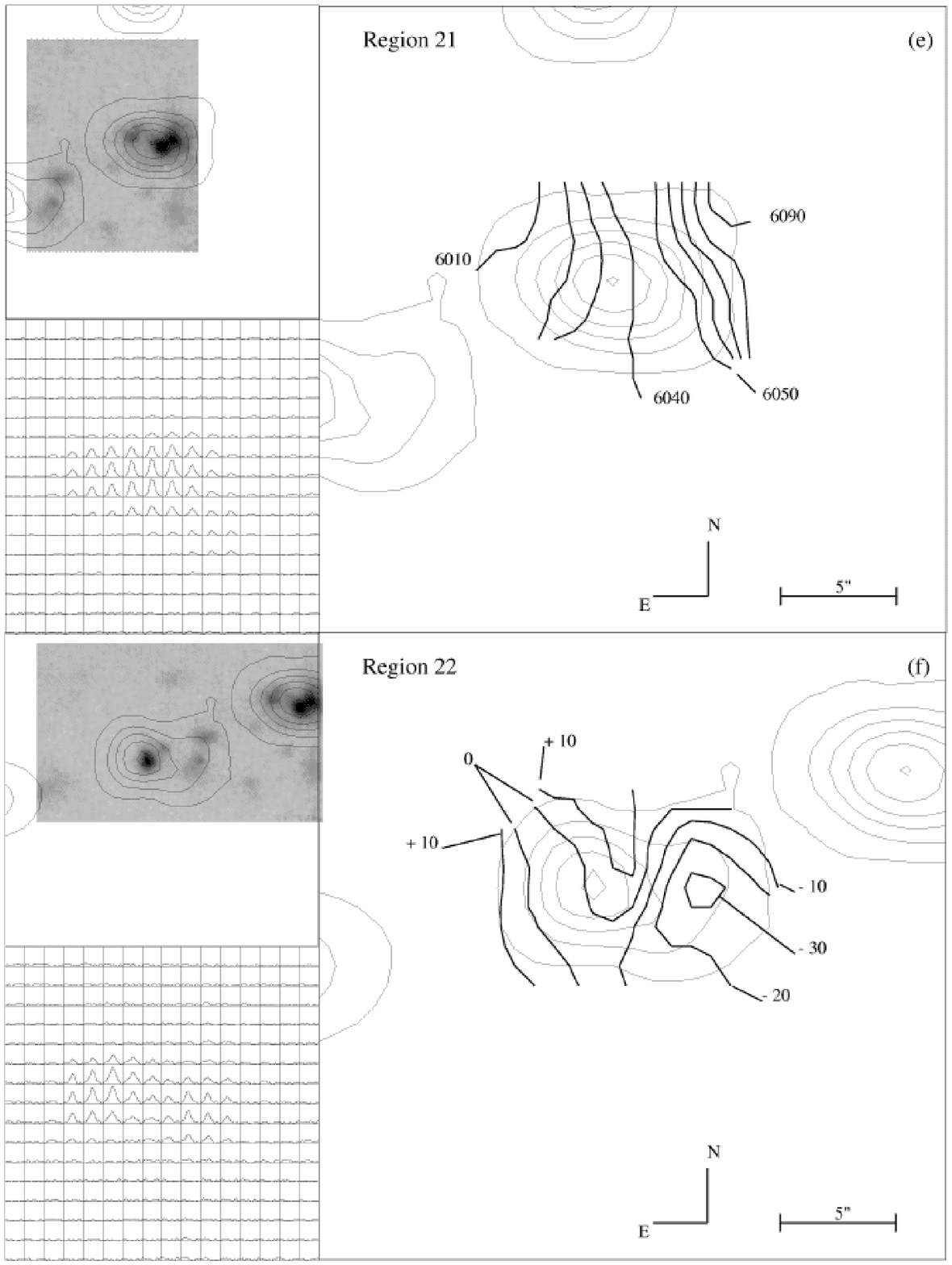}{14cm}{0}{77}{77}{-230}{-35}

\caption{}
\end{figure*}

\newpage

%
%

\begin{figure*}
\figurenum{3g}

\plotfiddle{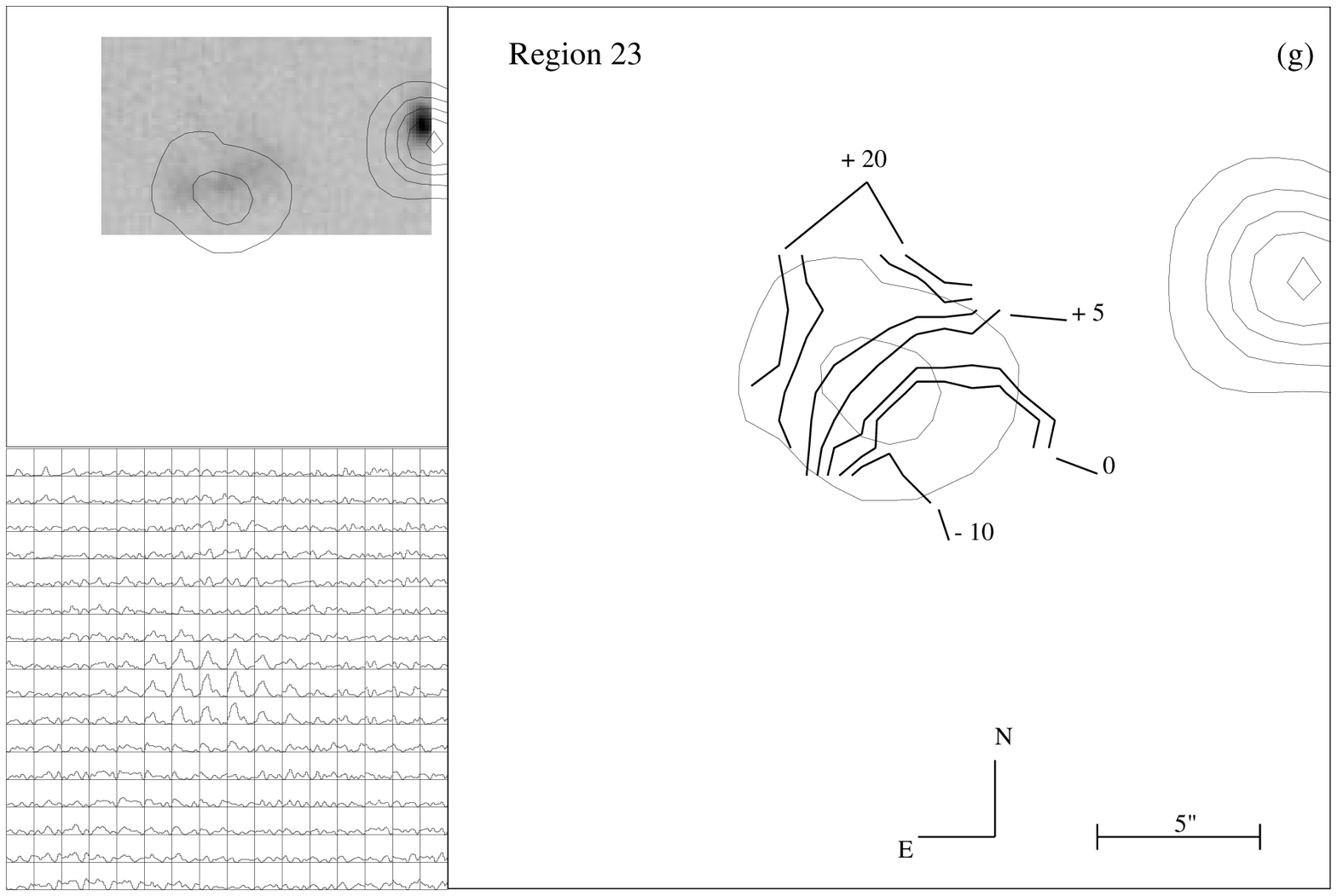}{14cm}{0}{77}{77}{-230}{-35}

\caption{}
\end{figure*}

\newpage

%
%

\begin{figure*}
\figurenum{4}

\plotfiddle{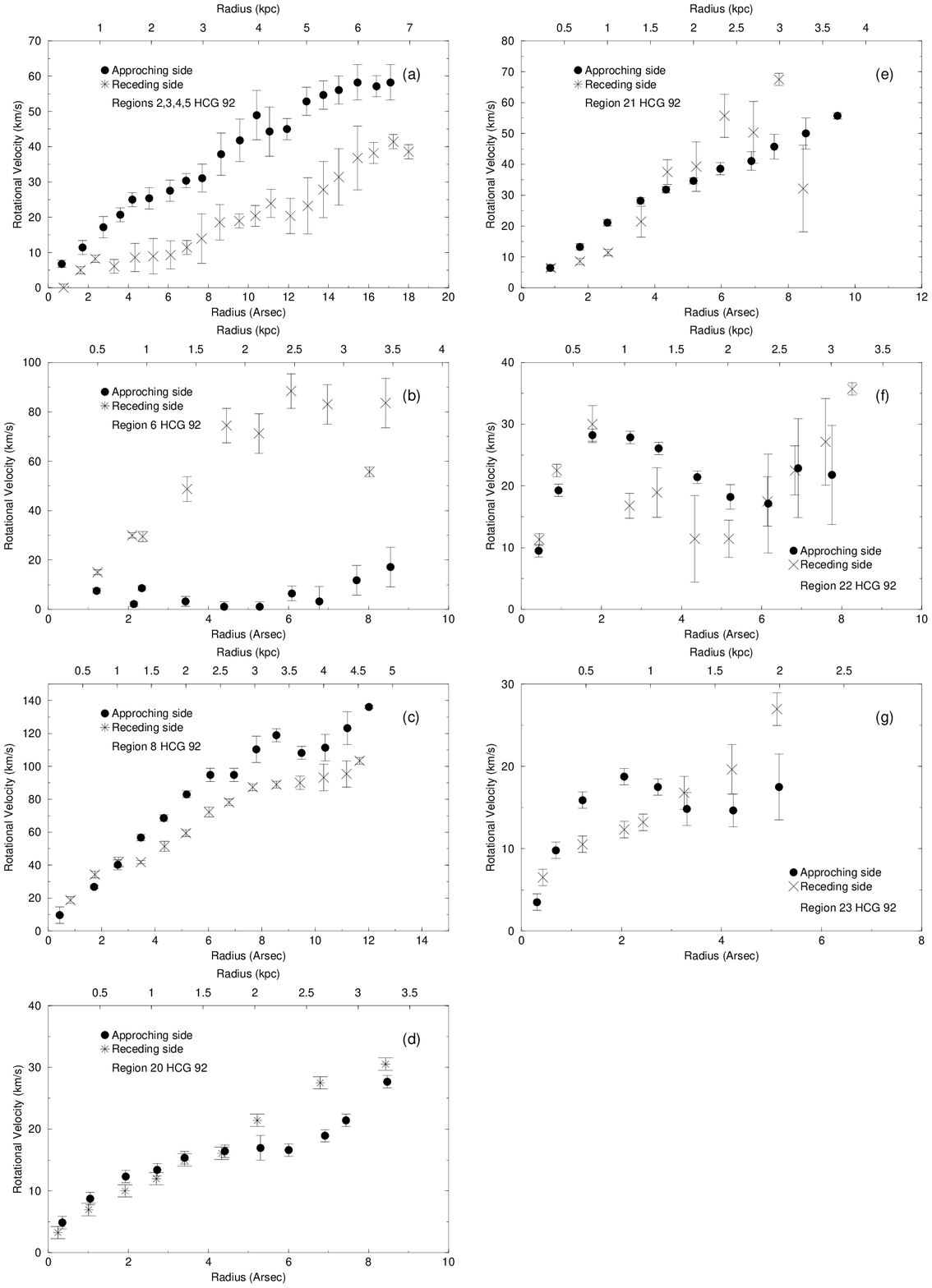}{14cm}{0}{80}{80}{-230}{-40}

\caption{Rotation curves of the seven dwarf candidate galaxies.
The parameters of the rotation curves are given in Table 3. }
\end{figure*}

\newpage

%
%

\begin{figure*}
\figurenum{5}

\plotfiddle{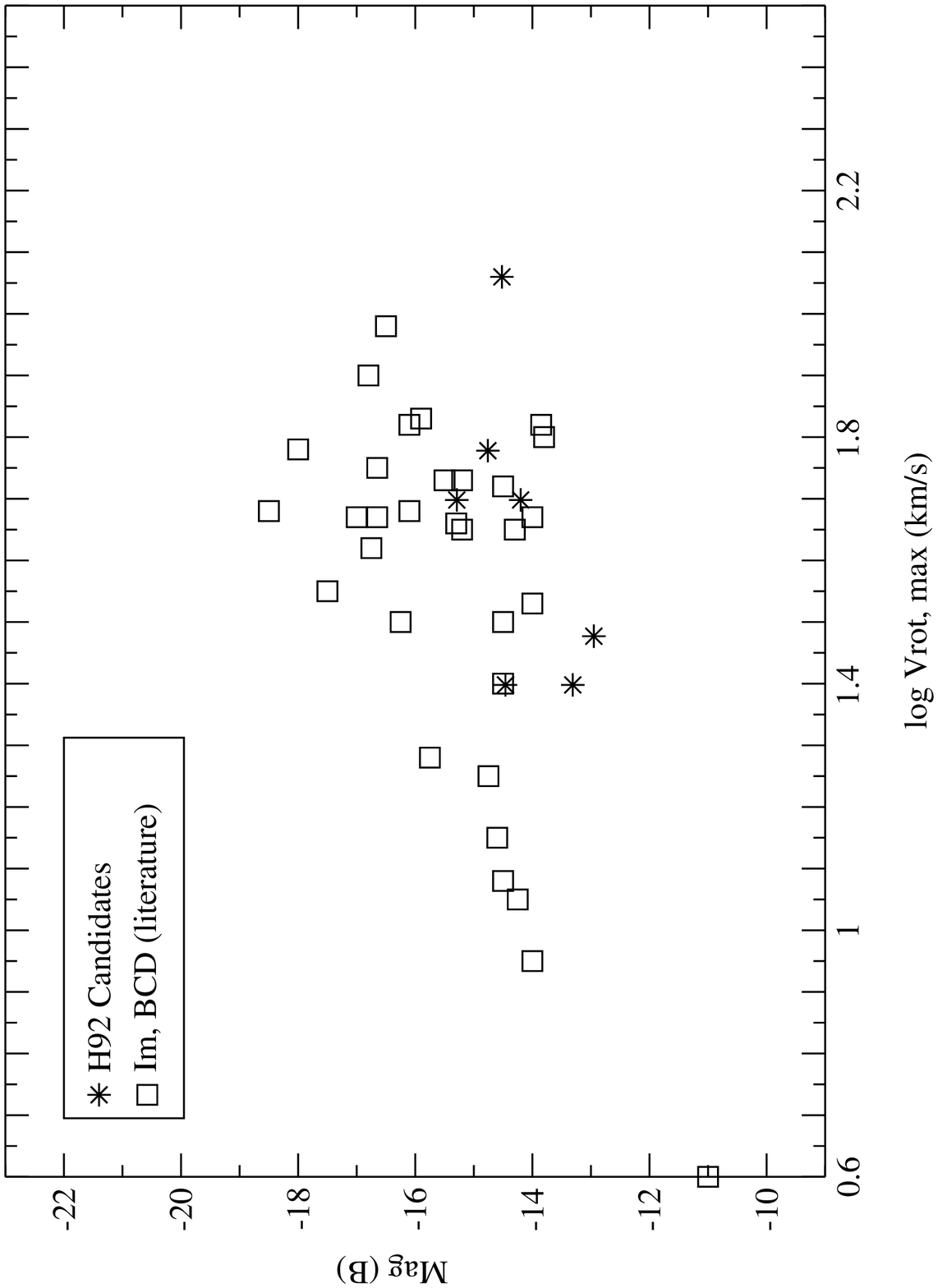}{14cm}{-90}{80}{80}{-300}{500}

\caption{
Plot showing the maximum rotation velocity
(corrected for the inclination) versus the absolute B magnitude, for both
the H92 tidal dwarf candidates and a compilation of Im and BCD galaxies
from Hunter et al. 2000.}
\end{figure*}

\newpage

%
%

\begin{figure*}
\figurenum{6}

\plotfiddle{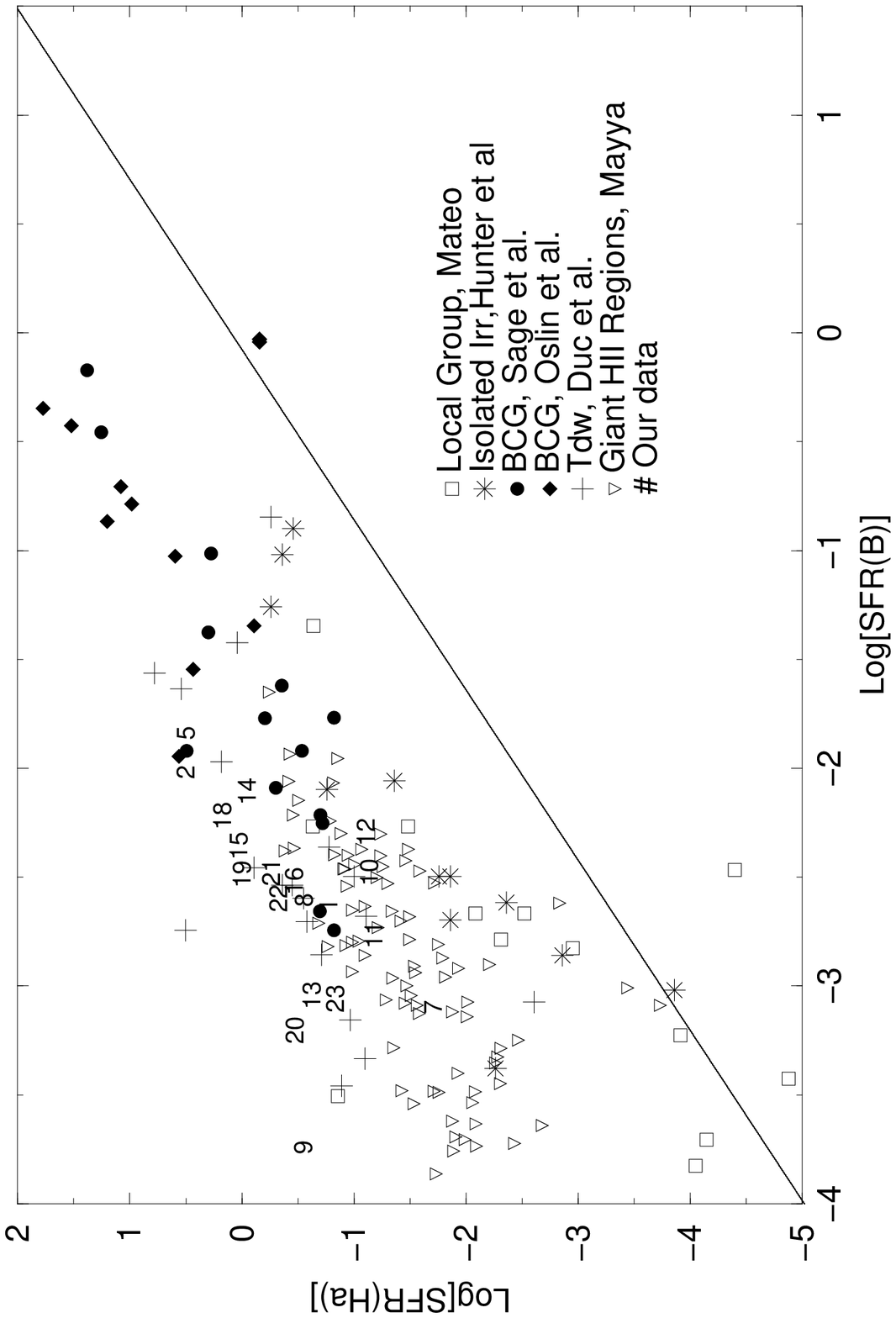}{14cm}{-90}{80}{80}{-300}{600}

\caption{A plot of the current vs recent star
formation rates for several different classes of emission line galaxies.
The line corresponds to the locus where both measurements of the star-formation rates are equal.}

\end{figure*}

\end{document}